\documentclass[fleqn,10pt]{wlscirep}
\usepackage[utf8]{inputenc}
\usepackage[T1]{fontenc}
\usepackage{bm}
\usepackage{lineno}
\usepackage{graphicx}
\usepackage{subcaption}

\title{Particle Physics and Gravitational Waves 
as complementary windows 
on the Universe}

\author[1,2*]{Steven D. Bass}
\author[3]{Laura Baudis}
\author[4]{Gianfranco Bertone}
\author[5,6]{Oliver Buchmueller}
\author[7]{Babette Döbrich}
\author[8,9]{Reinhard Genzel}
\author[10]{Anne M. Green}
\author[11]{Klaus Helbing}
\author[12,13,14]{Mich\`ele Heurs}
\author[15]{Karl Jakobs}
\author[16]{Markus Klute}
\author[4,13,14,17]{Samaya Nissanke}
\author[18,19,20]{Hiranya Peiris}
\author[21,22]{Albino Perego}
\author[23]{Stefan Pokorski}
\author[24]{Matthias Schott}
\author[21]{Stefano Vitale}
\author[25,26]{Georg Weiglein}
\author[8,27]{Jochen Weller}
\affil[1]{Kitzb\"uhel Centre for Physics, Kitzb\"uhel, Austria}
\affil[2]{
Jagiellonian University,
Marian Smoluchowski Institute of Physics,  Krak\'ow, Poland}
\affil[3]{Physik-Institut, University of Z\"urich, 
Z\"urich, Switzerland}
\affil[4]{Gravitation Astroparticle Physics Amsterdam (GRAPPA),
University of Amsterdam,  
The Netherlands}
\affil[5]{High Energy Physics Group, Blackett Laboratory, Imperial College, 
London, UK}
\affil[6]{Department of Physics, University of Oxford, 
UK}
\affil[7]{Max-Planck-Institut f\"ur Physik (Werner-Heisenberg-Institut),
Garching bei M\"unchen, Germany
}
\affil[8]{Max Planck Institute for Extraterrestrial Physics, 
Garching, Germany}
\affil[9]{Departments of Physics \& Astronomy, 
University of California, Berkeley, USA}
\affil[10]{School of Physics \& Astronomy, University of Nottingham, Nottingham, UK}
\affil[11]{Department of Physics, University of Wuppertal, 
Wuppertal, Germany}
\affil[12]{Institute for Gravitational Physics, Leibniz University Hannover, 
Hannover, Germany}
\affil[13]{Deutsches Zentrum für Astrophysik, 
Görlitz, Germany}
\affil[14]{
Deutsches Elektronen-Synchrotron DESY, 
Zeuthen, Germany}
\affil[15]{Physikalisches Institut, University of Freiburg,
Freiburg, Germany}
\affil[16]{Institute for Experimental Particle Physics (ETP), Karlsruhe Institute of Technology (KIT), Karlsruhe, Germany}
\affil[17]{Institut f{\"u}r Physik und Astronomie, Universit{\"a}t Potsdam, 
Potsdam, Germany
}
\affil[18]{Institute of Astronomy and Kavli Institute for Cosmology, University of Cambridge, 
Cambridge, UK}
\affil[19]{
Cavendish Laboratory, Department of Physics, University of Cambridge, Cambridge, UK}
\affil[20]{
The Oskar Klein Centre, Department of Physics, Stockholm University, Stockholm, Sweden}
\affil[21]{
Dipartimento di Fisica, Universit\`a di Trento, Trento, Italy}
\affil[22]{
INFN-TIFPA, Trento Institute for Fundamental Physics and Applications, Trento, Italy}
\affil[23]{Institute of Theoretical Physics, Faculty of Physics, University of Warsaw,
Warsaw, Poland}
\affil[24]{Physikalisches Institut, 
Rheinische Friedrich-Wilhelms-Universit\"at Bonn, 
Bonn,
Germany}
\affil[25]{Deutsches Elektronen-Synchrotron DESY,
Hamburg, Germany}
\affil[26]{
Institut für Theoretische Physik, Universität Hamburg,
Hamburg, Germany}
\affil[27]{Universit\"ats-Sternwarte M\"unchen, Fakult\"at f\"ur Physik, Ludwig-Maximilians-Universit\"at,
M\"unchen, Germany} 

\affil[*]{e-mail: Steven.Bass@cern.ch}

\begin{abstract}
Particle physics and 
gravitational waves provide  complementary probes of the deep structure of the Universe.
Gravitational waves from the mergers of neutron stars and black holes are sensitive to the structure of dense quark matter and to different dark matter scenarios. 
Measurements of stochastic gravitational waves backgrounds can teach us about possible first order phase transitions in the early Universe, including providing sensitivity to the TeV scale  which is of key interest to future particle collider experiments.  
Gravitational waves measurements will also give new probes of the evolution and expansion of the Universe, complementary to measurements 
with electromagnetic radiation. This Perspectives article explores the physics synergies between the science opportunities provided by next generation gravitational waves measurements and particle physics experiments. Gravitational waves can also probe deep into the early Universe reaching physics much above possible collider energies if the signals can be detected.
\end{abstract}

\begin{document}

\flushbottom
\maketitle

\thispagestyle{empty}

\section{Introduction}

Particle physics and 
gravitational waves (GWs) provide  complementary probes of the deep structure of the Universe. 
This is especially so with the next generation of experiments that will soon come on line in the next decade.
GWs emitted from the mergers of neutron star (NS) binaries tell us about the behaviour of dense matter and the phase diagram of Quantum Chromodynamics  (QCD). 
GWs from black hole (BH) binaries will shed new insight into dark matter (DM) scenarios.
Since GWs are only very weakly interacting, they can reach us from processes that happened in the early Universe before the cosmic microwave background (CMB). 
If the signal can be detected, 
they carry information about possible first order phase transitions in the early Universe, topological defects and even the inflationary era. 
Inspiral and merger observables will also tell us about the expansion of the Universe with  independent measurements of the Hubble parameter and dark energy (DE).
The NS measurements are complementary to future laboratory studies of the QCD phase diagram and dense matter at GSI/FAIR.
Looking for evidence of 
TeV scale phase transitions will be complementary to particle physics studies at the high luminosity upgrade of the LHC at CERN (HL-LHC) and possible future colliders.
DM might consist of new particles and/or be associated with BHs with both scenarios giving signals in future GW experiments.
Possible extra scalar fields could play an important role in the expansion and evolution of the Universe.
The energy scale of possible new physics beyond the Standard Model (SM) is so far not known from present experiments. 
One needs the widest possible search strategy.
GWs, if detected, can probe scales much above colliders as well as complementing the results of present and next generation laboratory experiments. 
In this Perspectives article we highlight synergies and key observables where the particle physics and GWs communities can work together to optimise the science output in our quest for deeper understanding of particle physics and its connections to cosmology.
At the interface 
there is an exciting programme of precision and discovery science for the next decade and beyond!

Particle physics today is characterised by 
the incredible success of the Standard Model. 
The Higgs boson discovered at CERN in 2012~\cite{ATLAS:2012yve,CMS:2012qbp}  behaves very SM like~\cite{Bass:2021acr,Jakobs:2023fxh}. Within the SM it completes the particle spectrum. 
Despite the success of the SM, 
there are still many open puzzles requiring deeper understanding. 
These include 
neutrinos 
(their tiny masses, might they be their own antiparticles and possible CP violation), 
families (why are there three?), 
baryogenesis (where does the matter-antimatter asymmetry in the Universe come from?), 
the non SM dark matter implied by astrophysics  (that clumps like normal SM matter under gravitation and is required by astrophysics to make up 27\% of the energy budget of the Universe) 
and dark energy driving the Universe's accelerating expansion (and making up 68\% of the energy budget) plus the physics 
associated with primordial inflation. 
All these are topics of vigorous 
investigation.
We would like to understand 
the origin of the 
gauge symmetries that determine particle dynamics of the SM 
plus the phase diagram of the SM including the physics underlying Higgs phenomena in electroweak interactions and the phases of hot and dense matter in QCD.
How does the picture of quark and gluon confinement in the infrared and perturbative QCD with asymptotic freedom in the ultraviolet generalise to finite densities and temperature?
Before neutrino masses, the minimal SM contains 18 parameters.
Of these, 15 are in the Higgs sector and 3 are the gauge couplings. 
How should we understand the underlying physics giving rise to this structure?
The structure of quark and gluon-made matter at NS densities 
will be investigated at GSI/FAIR in the 2030s.
At high energies 
a present prime push is for precision studies of the few TeV scale
with focus on Higgs dynamics 
and a view to maximally  constraining the SM
and 
hopefully new discoveries.
An important ingredient will be a precision measurement of the Higgs self-coupling which is essential in determining the shape of the Higgs potential and the stability of the SM vacuum.
The initial next step at the high energy frontier is the 
HL-LHC which is planned to run in 2030-2041.
This should be followed by a new 
$e^+ e^-$ collider as a dedicated Higgs factory. 
The physics details and science vision are given in the Physics Briefing Book~\cite{deBlas:2944678} of the 2026 update of the  
European Strategy for Particle Physics.
Prime collider options under consideration include the circular FCC-ee project at CERN or a future linear collider.
Various scenarios for new physics beyond the Standard Model give rise to signals in both particle physics and GWs observables
inspiring new synergies to explore.
For example, 
in models with extra Higgs states
the value of the Higgs self coupling 
is closely linked to the possibility of an early Universe first order phase transition that could be essential for explaining the observed baryon asymmetry and might be observed in GW measurements with the Laser Interferometer Space Antenna (LISA) experiment of the European Space Agency (ESA).
GW signals are expected for 
particle DM overdensities near BHs and from primordial BHs (PBHs).

\begin{figure}[t!]  
\centerline
{\includegraphics[width=0.90\textwidth]
{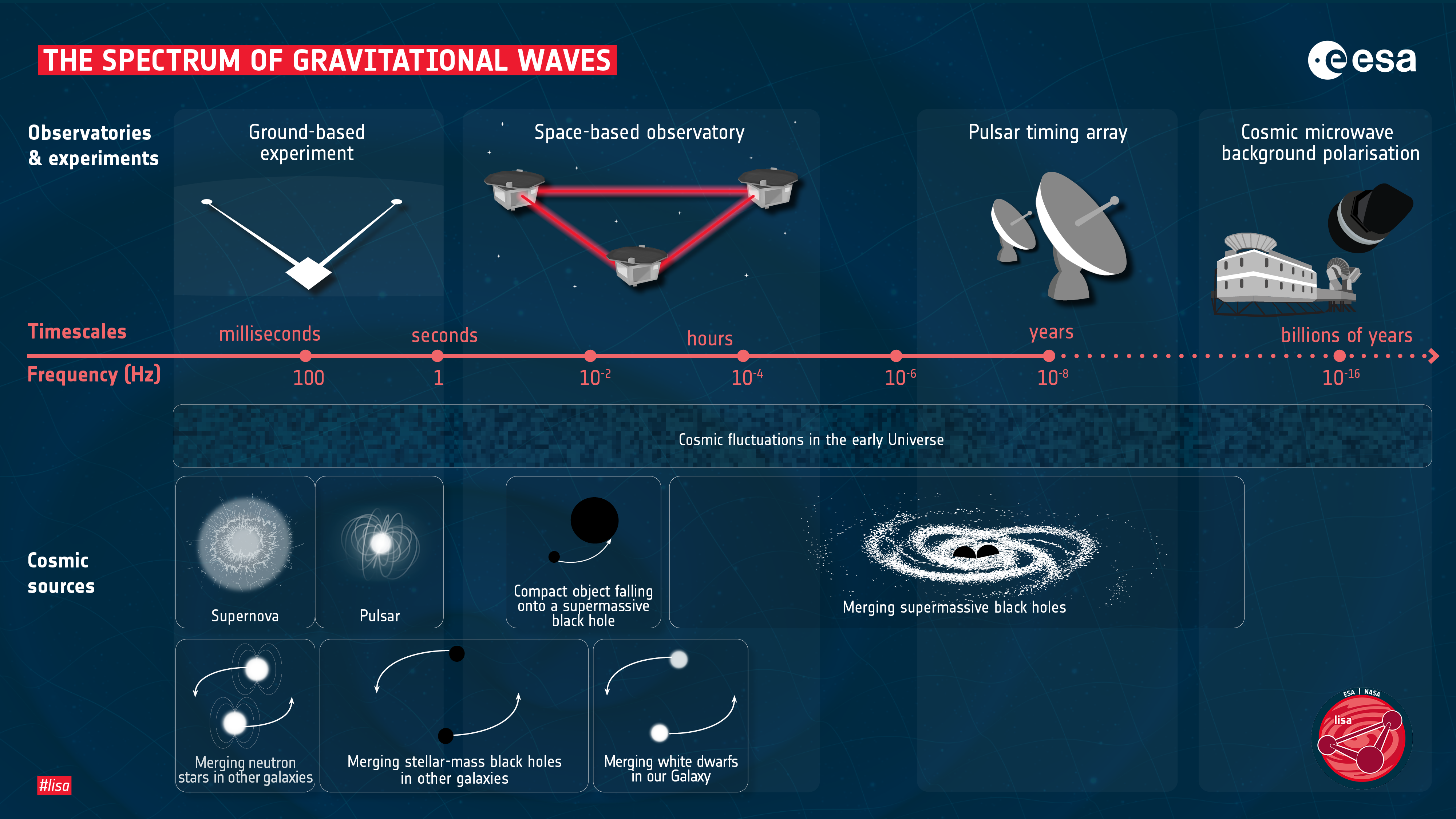}
}
\caption{ 
The gravitational waves spectrum showing phenomena that appear at different frequencies and the 
different types of GW detectors.
Figure credit: ESA.
}%
\label{fig:lisa}
\end{figure}

Gravitational waves, 
„ripples in spacetime“, 
are a prediction of General Relativity.
So far, GWs from mergers of compact objects, i.e. black holes 
and neutron stars, 
have been detected.
Since the first direct GW detection event (GW150914) in 2015, many more than 200 binary merger events \cite{LIGOScientific:2025slb,BHweb} and one multimessenger event from a binary NS merger event with accompanying electromagnetic observations (including a $\gamma-$ray burst and kilonova) 
\cite{LIGOScientific:2017vwq,LIGOScientific:2017zic}
have been recorded.
These events are characterised by perturbations of spacetime that propagate out from the source event. All GW events so far have been detected with kilometer-scale laser interferometers.
A passing GW periodically deforms spacetime differentially in the interferometer arms perpendicular to the GW's direction of travel, so that the interference pattern at the GW detector output changes.
The GWs cause a time-dependent dimensionless strain $h = \delta L/L$ where $L$ is the interferometer arm length (the distance between two mirrors, so-called „test masses“) and $\delta L$ is the GW-induced arm-length change over the baseline $L$.
For exact triangulation of GW sources, a worldwide network of GW detectors is required. The current second generation of GW detectors (aLIGO \cite{LIGOScientific:2014qfs}, Virgo \cite{VIRGO:2014yos}, and KAGRA \cite{KAGRA:2018plz}, short LVK) with armlengths of 3 to 4\,km are able to detect signals corresponding to a mirror displacement of less than 1/10000 times the size of the proton, or $10^{-19}$m.
To reach this extreme sensitivity, techniques employing non-classical (so-called „squeezed“) light and quantum sensors are required \cite{Heurs:2018wsu}. In addition to such transient merger event sources, stochastic GW backgrounds (SGWBs) are postulated.
A few standard deviations signal at GW frequencies $\sim 10^{-8}$ Hz 
has been observed in pulsar timing arrays \cite{NANOGrav:2023gor,EPTA:2023fyk,Reardon:2023gzh,Xu:2023wog},
which look for correlated signatures of pulse arrival times on Earth.
Here, a passing long-wavelength GW periodically changes the frequency of the signal from millisecond pulsars, and this is observed in the experiments.
This SGWB can be explained in terms of supermassive BH binaries at the centre of galaxies, though early Universe processes at the QCD scale cannot be excluded~\cite{Caprini:2024lxj}.
The GW detector network of the future will include spaceborne missions (e.g. LISA)
and will cover a larger range of detection frequencies, specifically including the milliHerz (mHz) range -- see Fig.~\ref{fig:lisa}. 
This will drive GW detection to a new level of precision and enable a plethora of new discovery science.
A GW Roadmap for the 2020s to 2030s is given in Ref.~\cite{Bailes:2021tot}. Besides mergers, future GW measurements will be sensitive to SGWBs, including from possible first order phase transitions in the early Universe, to ideas about DM and to
various models of inflation plus Lorentz invariance, the expansion rate of the Universe and DE.

Launching in 2035, the 
LISA mission will detect GWs 
with a frequency range between 10 $\mu$Hz and 1 Hz~\cite{LISA:2024hlh,Caprini:2025mfr}.
It will be sensitive to mergers involving 
supermassive BHs (SMBHs), 
typically weighing $10^5 - 10^9$ solar masses at the centres of galaxies, 
as well as to
extreme mass ratio inspirals (EMRIs), binary BH mergers with large mass asymmetry 
and on the order of ten thousand of resolvable Galactic  ultra-compact binaries (the majority being white dwarf binaries) \cite{LISA:2024hlh}.
The LISA sensitivity range also overlaps with GWs from any TeV scale first order phase transitions in the early Universe \cite{LISA:2024hlh,Caprini:2025mfr}.
LISA will form an equilateral triangle consisting of a laser-interferometric 
GW detector of 
three satellites 2.5M km apart and about 50M km from Earth.
Heliocentric orbits allow a quasi-stationary million km wide constellation. 
LISA will enable the detection 
of GWs with much longer wavelengths than ground 
based detectors.
Gravity-gradient noise on Earth prohibits 
GW detection at the low frequencies of LISA.
The LISA mission will have an initial lifetime of 4.5 years.
LISA will use 
free-falling test masses 
which have no contact with the spacecraft they will reside in.
Acceleration noise will be controlled so that the free test masses accelerate only because of GWs.
The technology has been very successfully tested in the LISA Pathfinder mission \cite{LISAPathfinder:2024ucp}. 
The LISA science programme is described in 
Refs.~\cite{LISA:2022kgy,LISACosmologyWorkingGroup:2022jok,LISA:2022yao}.

Looking forward, the next (third) generation of ground-based laser-interferometric GW detectors, e.g. 
the Einstein Telescope (ET) in Europe and the Cosmic
Explorer (CE) in the United States, are currently under active development. 
They will increase the sensitivity to GWs by more than an order of magnitude in the frequency range 
between a few Hz and 2000 Hz 
above the projected 2028-2030 sensitivities of the present 
second generation GW detectors.
They should detect all stellar-mass BH mergers in the observable Universe 
reaching back to the period before the formation of stars in the early Universe. 
There is a rich scientific programme with close connections to particle physics, astrophysics and cosmology including 
key observables connected to QCD, DE and DM. 
The science programme for the ET is given in Refs.~\cite{Branchesi:2023mws,Abac:2025saz,ET:2019dnz} 
and for the CE in 
Refs.~\cite{Evans:2021gyd,Evans:2023euw}. 
The aim is for GW detections in the late 2030s and beyond with a running lifetime of about 50 years. 
In parallel, 
there is also a vigorous R\&D program aimed at using new atom interferometer technologies to measure GWs plus to search for possible ultralight DM candidates.
Within the recently established Terrestrial Very Long Baseline Atom Interferometry (TVLBAI) collaboration~\cite{TVLBAISummary1, TVLBAISummary2}, first-generation vertical atom-interferometer experiments of length ${\cal O}(10)$ m have been constructed in Hannover~\cite{schlippert2020matter}, at Stanford~\cite{Overstreet:2021hea} and in Wuhan~\cite{zhou2011development}, a Technical Design Report for a 10 m experiment at Oxford has been completed~\cite{Bongs:2025rqe}, and ${\cal O}(100)$ m vertical and horizontal atom-interferometer experiments are under construction or proposed at Fermilab (MAGIS-100)\cite{MAGIS-100:2021etm}, in France (MIGA)\cite{Canuel:2017rrp}, and at CERN (AICE)~\cite{Baynham:2025pzm}.
High-frequency gravitational waves (HFGWs) detectors are being investigated with the aim to reach sensitivity up to about 1 GHz or even THz to look for new physics signals from the early Universe \cite{Aggarwal:2020olq,Berlin:2021txa,Franciolini:2022htd}.
Inflation era GW effects should also be manifest in polarisation observables in the CMB.
There are ideas to access the mid-frequency region between LVK and LISA 
using cold-atom interferometry \cite{Badurina:2021rgt} or a new laser interferometer in space~\cite{Kawamura:2020pcg}.

In the next sections we highlight key physics topics and observables where particle physics and GWs measurements will provide essential and complementary information.
There are exciting prospects for future collaboration both with physics synergies and new technology developments. 
If we see something new in a particle physics or GW measurement, what might its signal be in the other? 
How might the different measurements complement each other? 
If one finds a new particle, how might it feature in the early Universe with first order phase transitions or in  clouds in space near BHs, each with GW signals. 
Likewise, if we see something new in the GW spectrum, what might it mean for particle physics and does it have a particle physics correspondence?

\section{Black hole and neutron star binaries 
probing QCD, dark matter and the expanding Universe}

The experimental discovery of GWs by LIGO-Virgo in 2015 involved the BH merger GW150914 of two BHs each weighing 29 and 35 
solar masses with 3 solar masses, 
$\sim 5\%$ of the initial mass, 
released as GWs \cite{LIGOScientific:2016aoc}.
This discovery event 
was seen in the GW 
frequency range 35-250 Hz. 
The GW luminosity at merger was larger than that emitted by all stars
in the visible Universe.
Today the more than 200 BH mergers recorded by LVK 
tell us much about the distribution of BHs in the redshift region $z \lesssim 1$.
Future GW measurements will probe the origin and evolution of BHs deep into the early Universe 
plus also 
BH environments as a test of particle DM scenarios.
Recent results from GWs and from telescopes challenge our understanding about the origins and properties of BHs beyond the BHs with stellar origins. 
GW observations from LVK have also revealed the mergers of new types of BHs 
with masses between about 60 and 130 solar masses,  
beyond the theoretically expected range from stellar collapse processes
\cite{LIGOScientific:2020iuh,LIGOScientific:2025rsn}. 
Might these mergers 
be secondary merger events or might they involve a 
new species of 
intermediate mass black holes (IMBHs) 
too heavy to have formed from stellar origins and lighter than SMBHs, and formed 
with some other origin in the early Universe?
SMBHs are observed to exist at the centres of galaxies~\cite{Genzel:2021lmp}. 
One finds phenomenological correlations between their  properties and the properties of their host galaxies including their DM halos 
\cite{Ferrarese:2002ct,Ferrarese:2004qr,Ferrarese:2006fd,Genzel:2024vou}.
The Hubble and JWST 
optical telescopes reveal that SMBHs were already there in the early Universe at high redshifts. 
For example, 
the Capers-lrd-z9 galaxy is seen at redshift $z=9.288$
when the Universe was just 500 million years old 
with an estimated SMBH mass 
between $4.5 \times 10^6$ and $3.2 \times 10^8$ solar masses 
\cite{2025ApJ...989L...7T}. 
How did the SMBHs form and how did they get so big so quickly?
Also, the newly discovered SMBH 
QSO1 with mass 50 000 000 solar masses at redshift $z=7.04$ appears as a SMBH without a surrounding galaxy.
It has been interpreted as a possible primordial BH candidate \cite{Maiolino:2025tih}.
These observations challenge our ideas about evolution of structure in the Universe. 
Further, 
one finds the intriguing observation that SMBHs are spinning very fast today \cite{Reynolds:2013rva}
whereas GW measurements from LVK give  slower spinning BHs in the mass range detected \cite{McKernan:2023xio}.
Sgr A* at the centre of the Milky Way is spinning with 60\% of the maximum BH angular momentum 
\cite{Daly:2023axh} and the SMBH M87 at 80-99.8\% \cite{Drew:2025euq}. 
Future GW measurements will test that spinning BHs really satisfy the Kerr metric of General Relativity.
Measurements of early galaxies at the peak of galaxy formation, redshifts $z \sim 10$ (or 10 billion years back) suggest that these galaxies may have contained less DM as determined by their rotation curves
\cite{Genzel:2017jgd}.
The origin of 
IMBHs and possible connection to DM and also SMBHs including possible PBHs, their origins, seeds and evolution are important topics of investigation \cite{Genzel:2024vou,Volonteri:2021sfo}. 
LISA and the JWST will be complementary in their redshift reach with LISA sensitive to SMBH mergers up to $z \lesssim 15$. 
Binary systems in their gradual inspire phase are approximately Keplerian.
At zeroth order, 
the frequency of the GWs emitted in their mergers is given by 
\begin{equation}
   f_{\rm GW} 
   = 2 f_{\rm Kepler} \sim r^{-3/2} M^{-1} 
\end{equation} 
where $M$ is the total mass of the binary and $r$ the distance between the two merging objects (or the size of the binary).
For a given binary of total mass $M$, at merger the heavier the merging BHs means the smaller the frequencies and the larger the wavelengths.
SMBH mergers require the low frequency sensitivity of LISA. 
With smaller frequencies one needs a larger baseline to see SMBH mergers.
SMBH mergers involve a huge release of energy in GWs. 
For example, 
for an equal mass merger if SMBHs of $10^6$ solar masses 
will release 
$\sim 10^5$ solar masses 
as GWs corresponding to 5\% of the total mass involved.  
LISA will also 
measure extreme mass ratio mergers up to $z \lesssim 1$.
Measurements of the evolution of 
the SMBH systems will be 
complementarity to studies of SMBH/galaxy systems with optical telescope measurements from the JWST space telescope and the future European Large Telescope in Chile.
For stellar-mass BHs and IMBHs
the ET/CE should 
be able to see $\sim 10^5$ mergers per year compared to the about 200 so far combined from LVK.
They be able to probe the entire cosmic history of 
BHs with masses between about 10 and 100 solar masses 
reaching up to redshifts 
$z \lesssim 20$, and   
for the lower part of this mass range up to $z\sim 100$.
These measurements will push the sensitivity to stellar mass BHs into the so called dark ages preceding the birth of the first stars. 
Mergers from before this time would involve BHs of primordial origin.
Might DM be residing, at least part, in BHs?
There are 
current gaps in theoretical understanding beyond the current electromagnetic  redshift frontier $z \sim 15$ and the ET/CE horizon $z \sim 100$. 
The challenge to understand this early Universe dynamics calls for new theoretical ideas and modelling efforts in this territory, 
especially in directions that go beyond phenomenology.

A wide array of astrophysical and cosmological observations provide strong evidence for the existence of dark matter, a non-luminous component that contributes about $85\%$ of the matter density of the Universe \cite{Baudis:PDG,Bertone:2004pz}. In the standard cosmological model, dark matter is stable, cold, and only interacts with Standard Model particles via gravity. Possible candidates include new elementary particles such as weakly interacting massive particles (WIMPs), axions and sterile neutrinos, as well as macroscopic candidates such as primordial black holes (PBHs).
A vigorous programme of laboratory and astroparticle 
experiments is underway.
Gravitational waves, collider experiments, and astroparticle searches probe complementary aspects of dark matter: its gravitational imprint, its particle interactions, and its cosmological abundance. Only their combination can disentangle particle properties from astrophysical systematics \cite{Bertone:2018krk}.

Collider searches provide a complementary and largely model-independent approach to the identification of particle dark matter by probing its production and interactions under controlled laboratory conditions \cite{DeRoeck:2024fjq}. At the LHC, dark-matter candidates are searched for via events with large missing transverse momentum recoiling against visible Standard Model objects such as jets, photons, or electroweak gauge bosons. These so-called mono-X signatures test scenarios in which dark matter couples to fermions or bosons through heavy mediators, and are sensitive to a wide range of masses and interaction structures. In parallel, precision measurements of the Higgs boson offer a powerful probe of Higgs-portal models, in particular through constraints on the Higgs invisible decay width and exotic Higgs decays. Collider searches also target richer dark sectors, including extended Higgs sectors, supersymmetric particles, dark photons, and axion-like particles, which may evade direct detection due to suppressed couplings or non-standard cosmological histories. While no evidence for dark matter has yet been observed at colliders, the High-Luminosity LHC will substantially extend the sensitivity to mediator masses and couplings, and future lepton and hadron colliders would enable percent-level precision tests of Higgs portals and electroweak dark sectors. Any collider signal would play a crucial role in interpreting astrophysical or gravitational-wave observations by establishing the particle nature and interaction strengths of dark matter, thereby anchoring cosmological scenarios in laboratory measurements.

Direct detection experiments look for search for elastic or inelastic scatters of DM particles with atomic nuclei or with electrons in the atomic shells of various target materials. These interactions are classified as nuclear recoils (NRs), and electronic recoils (ERs), respectively. The main physical observable is a differential recoil spectrum (or a line feature for the absorption of, e.g.,  axions and dark photons), and its modelling relies on inputs from particle physics, astrophysics, atomic and nuclear physics, and more recently also material science. The experiments are designed to observe low-energy and rare signals which are induced by DM particle scatters in detectors operated deep underground. The observed signals are in the form of ionisation, scintillation or lattice vibrations, with most experiments detecting more than one signal, allowing one to distinguish between ERs and NRs. A three dimensional position resolution is required to define central  detector regions with low background rates due to the radioactivity of surrounding detector components. The ability to separate single- and multiple-scatters is used to reject a significant fraction of backgrounds, considering that a DM particle will scatter at most once in a given detector.  A variety of techniques are employed to search for the rare interactions expected from galactic DM particles. While most experiments have sensitivity to a wide range of particle masses, some technologies were optimised for light and heavy DM, respectively. There is an ongoing R\&D programme to observe the direction and sense of a nuclear recoil, correlated with the direction of the incoming WIMP.  While there is no sign for a DM-induced signal yet, ongoing and next-generation experiments will cover a much larger region of the allowed parameter space \cite{Baudis:2023pzu,Aprile:2024fkg}. The main challenges are to lower the energy thresholds further, to reduce and characterise the background sources and at the same time to increase the DM target masses of the detectors. While the hope is to discover a new, DM particle species, direct detection experiments will explore the experimentally accessible parameter space until the observed recoil rates will be dominated by astrophysical neutrinos.

For possible light mass pseudoscalar axions, 
there are several search strategies \cite{Dobrich:2025oso}.
One is through searches for rare decays, e.g., charged kaon decays $K^+ \to \pi^+ a$ where $a$ is the axion and $\pi^+$ is a charged pion  \cite{Goudzovski:2022vbt} with present limits given in Ref.~\cite{NA62:2025upx}. 
Alternatively, one may look for their conversion to photons in the
presence of a large magnetic field
or production of axions from two photon collisions with the axion passing through a wall and then decaying into two photons again in the experiments.
Recent LVK constraints on ultralight DM particles interacting with SM particles in the instruments are given in Ref.~\cite{LIGOScientific:2025ttj}.

\begin{figure}[t!]  
\centerline
{\includegraphics[width=0.50\textwidth]
{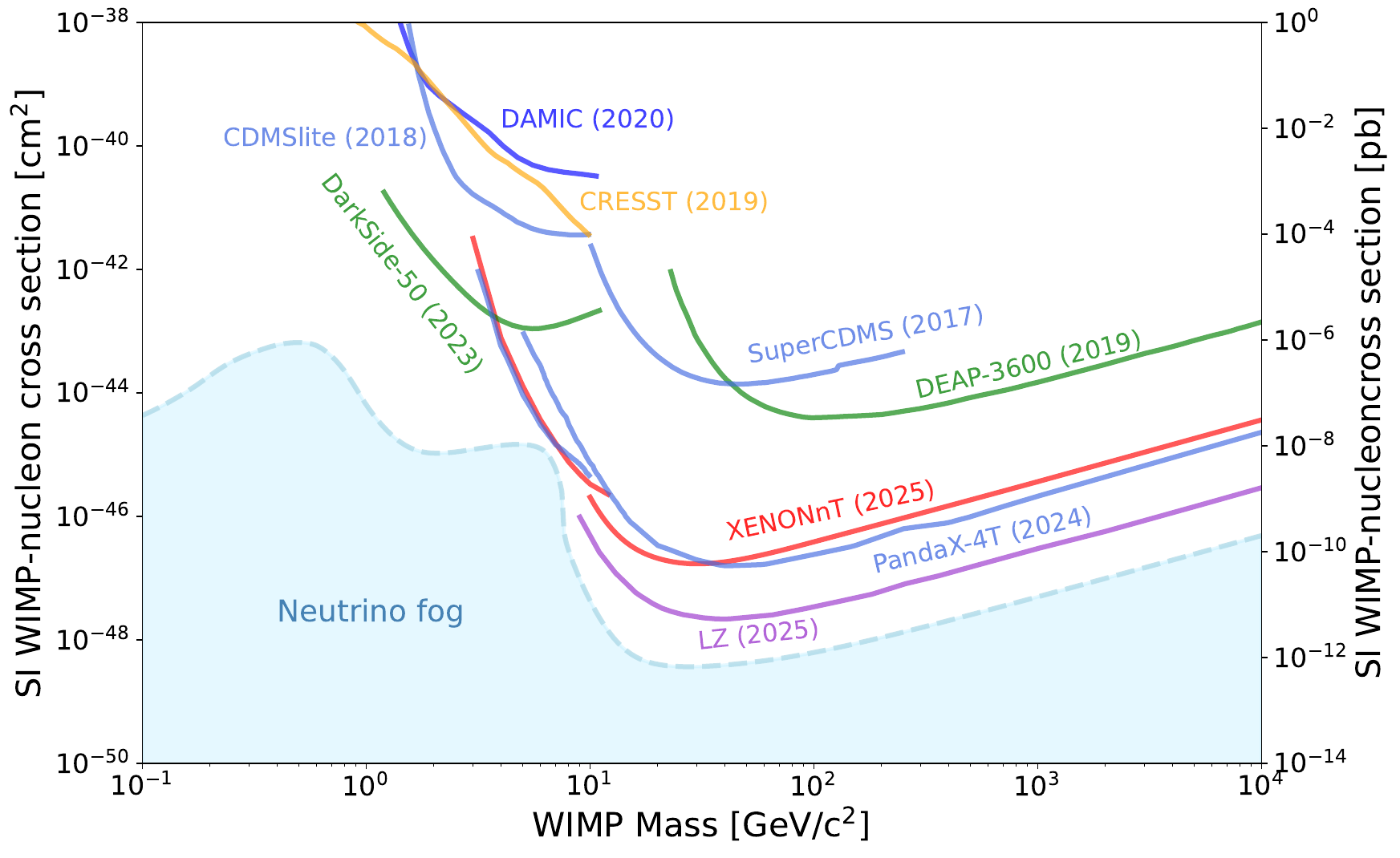} 
\hfill
\includegraphics[width=0.45\textwidth]
{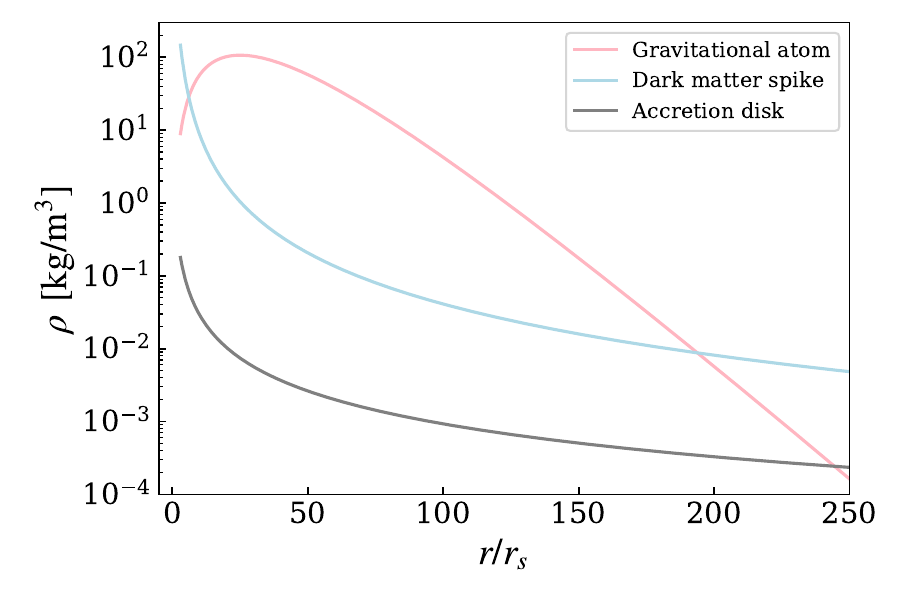}
}
\caption{
Left: Upper limits on the spin-independent (SI) WIMP-nucleon cross section as a function of DM mass from direct
detection experiments. The region where astrophysical neutrinos will start to be a limiting background is indicated in blue
(neutrino fog). Figure from Refs.~\cite{ParticleDataGroup:2024cfk,Baudis:2025yva}, updated (PDG2025).
Right: 
DM density profiles of an accretion disc, a WIMP dark matter spike and a gravitational atom, surrounding a $10^5$ solar masses black hole. For details, see Ref.~\cite{Cole:2022yzw}.
Here $r$ is the radial distance and $r_s$ is the Schwarzschild radius of the BH. The resulting DM  environment-induced dephasing in the gravitational waveform of intermediate- and extreme-mass-ratio inspirals is potentially detectable with future interferometers.
}
\label{fig:DM}
\end{figure}

Dark matter can be searched for indirectly, by looking for the Standard Model products of dark-matter annihilation or decay, in particular gamma rays, antimatter and neutrinos emitted from regions of enhanced dark-matter density such as the Galactic Centre, dwarf spheroidal galaxies and galaxy clusters. Combined Fermi-LAT observations of Milky Way dwarf spheroidals now provide some of the most stringent and robust limits on the annihilation cross-section of WIMP-like dark matter \cite{McDaniel:2023bju}, while forecasts for the Cherenkov Telescope Array indicate that next-generation observations of the inner halo and nearby dwarfs could further improve the sensitivity to multi-TeV dark matter by up to an order of magnitude relative to current gamma-ray constraints \cite{Lefranc:2015pza}.

Gravitational waves open new opportunities for dark matter searches, ranging from tests of macroscopic dark matter candidates such as primordial black holes and exotic compact objects, to probes of dark matter overdensities and ultralight boson clouds around compact binaries, and even to direct detection of dark matter fields through their couplings to interferometers~\cite{BertoneTait:2018,BertoneCroon:2020GWDM,Bertone:2024DMBHGW}. If dark matter consists of ultralight bosons with Compton wavelength comparable to the gravitational radius of a spinning black hole, superradiant instabilities can build up a ``gravitational atom'', whose resonant transitions and ionisation by a binary companion can dominate the energy losses and imprint a characteristic dephasing in the gravitational waveform~\cite{Bertone:2024DMBHGW,Tomaselli:2025jfo}. Cold dark matter can also be compressed into steep density ``spikes'' \cite{Gondolo:1999ef,Sadeghian:2013laa} or more extended ``mounds'' \cite{Bertone:2024wbn,Caiozzo:2025mye} around massive black holes, which modify the orbit and ensuing gravitational waveform of compact objects inspiralling within these overdensities \cite{Eda:2013gg,Kavanagh:2020cfn,Coogan:2021uqv,Cole:2022yzw,Bertone:2024DMBHGW}. The resulting environment-induced dephasing in the gravitational waveform of intermediate- and extreme-mass-ratio inspirals is potentially detectable with future interferometers, allowing not only the discovery of dark-matter structures around black holes but also few-percent-level constraints on their density profiles and a robust discrimination from vacuum general-relativistic waveforms~\cite{Bertone:2024DMBHGW}.

The existence of 
PBHs is an open topic. 
While QSO1 could potentially be a PBH, other observations exclude such massive PBHs making up all of the DM 
\cite{Green:2024bam,Green:2020jor}. 
If PBHs should comprise a significant fraction of the DM, 
theoretical ideas commonly favour asteroid size PBHs.
In this case one expects a scalar SGWB signal at LISA frequencies with might be looked for in future experiments \cite{Caprini:2025mfr}.
DM being a mixture of WIMPs and PBHs has (essentially) already been excluded due to the non-observation of gamma-rays from WIMP annihilation in ultracompact mini-halos around PBHs \cite{Lacki:2010zf}. 
There is also a proposal that BH "remnants" after Hawking evaporation in the early Universe might be DM candidates \cite{MacGibbon:1987my,Franciolini:2023osw}. 
PBH mergers are a target for future HFGWs measurements \cite{Aggarwal:2025noe}.

\begin{figure}[t!]  
\centerline
{\includegraphics[width=0.45\textwidth]
{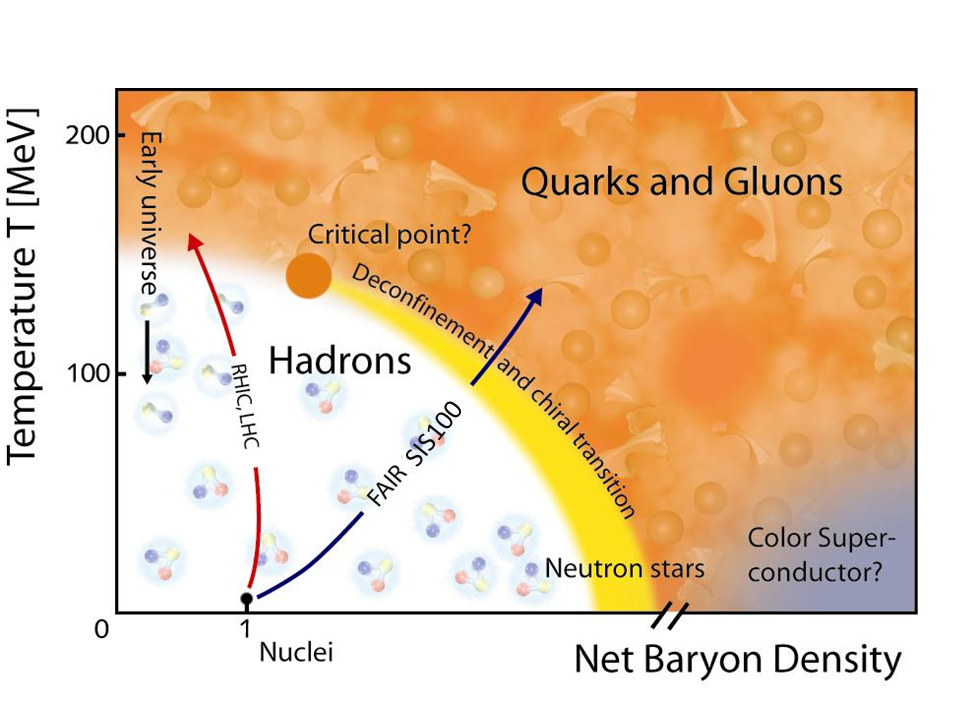}
\includegraphics[width=0.45\textwidth]
{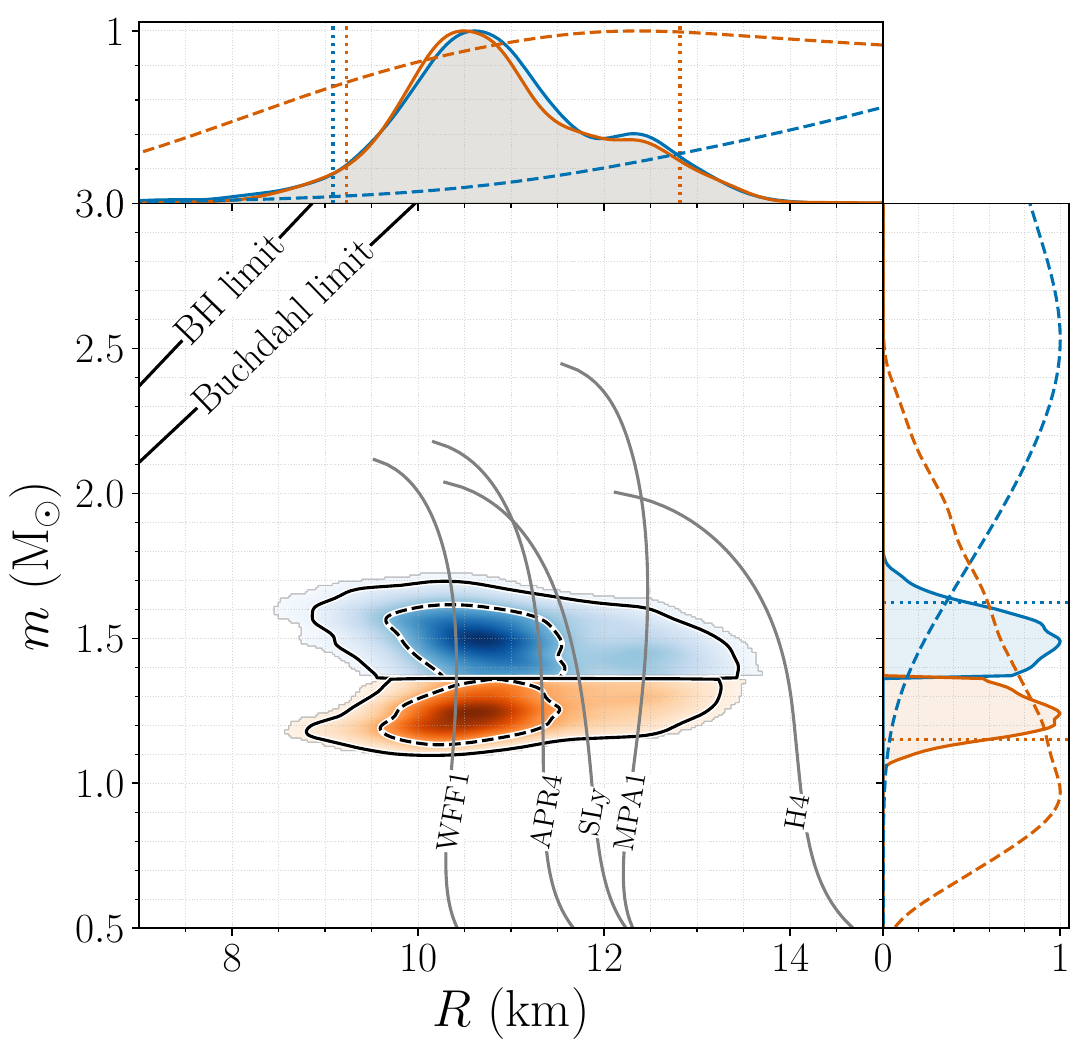}}
\caption{Left: 
The QCD phase diagram showing the domains of high density neutron stars and the physics to be explored in forthcoming experiments at GSI/FAIR as well as measurements with high energy heavy-ion collisions at the LHC and RHIC.
Figure credit: GSI. 
Right: Neutron star properties deduced from the NS merger event GW170817. 
The posterior for the mass $m$ and areal radius $R$ of each binary component that produced the merger,  
using an EoS-insensitive relations method. 
The top blue (bottom orange)
posterior corresponds to the heavier (lighter) NS and several model curves are shown for comparison. 
Figure from 
Ref.~\cite{LIGOScientific:2018cki}.
}
\label{fig:NS}
\end{figure}

Synergies between the GW and electromagnetic communities will provide essential new information about the neutron star equation of state (EoS) and about the expansion rate of the Universe.
GWs from NS mergers tell us
about the structure of 
QCD matter at large nuclear densities. In
its low energy limit, QCD features the appearance of baryons and pions as
relevant degrees of freedom, while high enough densities and temperatures
involve a possible phase transition to quark matter. 
The related QCD phase diagram~\cite{Stephanov:2004wx} is an important topic of exploration for experiments at the LHC (the ALICE experiment), RHIC and GSI/FAIR.
Key issues are the change between strong interaction degrees of freedom and the nature of any transitions at different temperatures and densities, whether first or second order or crossovers, plus the search for a predicted QCD critical point, 
see Fig.~\ref{fig:NS}. 
NSs are the densest, stable, known objects made of QCD matter with their cores reaching several times that of nuclear density
and
one of the end points of stellar evolution, with intermediate properties
between a white dwarf and a BH. NSs come with radius about 10 - 12 km
and masses up to about 2 - 2.3 solar masses.
During the inspiral of a neutron star binary, 
the leading correction 
due to finite size effects 
to the 
binary dynamics and GW emission
(with respect to point particle dynamics) 
comes from tidal interaction. The resulting inspiral signal and tidal phasing
can directly constrain regions in the NS EoS at around the maximum density
of the NSs in the binary, that for the fiducial binary neutron star (BNS) is around twice nuclear 
saturation density 
\cite{LIGOScientific:2018cki,Agathos:2019sah}. In contrast, BH mergers have no
tidal effects with BHs characterised just by their masses and spins. 
The neutron rich matter expelled by 
NS mergers is believed to be one of the main source of the r-process nucleosynthesis of heavy elements.
This involves the rapid capture of neutrons, captured on timescales faster than the $\beta-$decays of the resulting daughter nuclei.
The r-process is believed to be responsible for the origin of about half of the elements heavier than iron \cite{Siegel:2022upa}.

The multi-messenger 
NS merger event GW170817 
was observed by LIGO-Virgo in coincidence with a high energy $\gamma-$ray burst 
(detected by the Fermi and INTEGRAL satellites with arrival $\sim 1.7$ seconds after the GW peak) at a distance 40 Mpc away at redshift $z=0.0099$ \cite{LIGOScientific:2017vwq,LIGOScientific:2017zic}.
The signals arrival times 
tell us that the speed of GWs and light are the same, at least to within one part in $10^{15}$, as predicted by relativity and providing a mass bound on the graviton (if it exists). 
In the days and weeks 
following the GW170817 merger event a kilonova was observed. This is an isotropically expanding luminous transient of electromagnetic radiation emitted by the radioactive decay of r-process nuclei synthesised by and then ejected from the initial merger event.
With GW170817
the total initial mass was  $2.74^{+0.04}_{-0.01}$ solar masses. 
In the analysis of 
Ref.~\cite{LIGOScientific:2018cki} 
one finds a 90\% chance that the heavier NS had mass between 
1.36 and 1.62 solar masses and the lightest NS between 
1.15 and 1.36 solar masses, 
see Fig.~\ref{fig:NS}.
The total mass involved and 
the absence of magnetic effects in the kilonova and GRB emission suggest that the remnant was probably a BH, thus putting an upper bound on NS maximum masses. 
This mass limit also constrains the NS EoS with the result between maximum hardness (meaning just compressed neutrons) 
and softness 
(with smallest pressure for a given density).
In general, a hard EoS means a larger radius and bigger maximum mass for the NS \cite{Yunes:2022ldq}.
The GW results on the NS EoS are consistent with what is extracted from X-ray observations of NSs \cite{Miller:2019cac,Miller:2021qha}. 
One finds a jump from non relativistic to relativistic matter at between one and 2-3 times nuclear matter density with a transition to some new form of strongly coupled and conformal dense matter in the NS core \cite{Fujimoto:2022ohj}.
This matter might involve some combination of $\Delta$ resonances \cite{Metag:1992jh,Ribes:2019kno} 
and hyperons \cite{Tolos:2024xyi} 
or some new (colour superconducting) quarkyonic matter \cite{McLerran:2018hbz}.
Details of the transition are expected to be manifest in improved precision measurements of the waveforms. 
The postmerger is the phase with the largest GW luminosity: up to the equivalent of 0.1 solar masses are radiated over 
$\sim 10 - 20$ms 
\cite{Zappa:2017xba, 
Bernuzzi:2015opx}, with a spectrum characterized
by discrete features or peaks. The main one
is a broad peak
at frequencies 2-4 kHz, accessible to next generation GW detectors 
\cite{Hotokezaka:2013iia, 
Bauswein:2011tp,
Dietrich:2016hky}.
For a given total mass $M$ of the binary, this frequency 
correlates with the NS radius
or tidal deformability thorugh quasi-universal (i.e. EoS independent)
relations 
\cite{Breschi:2021xrx,Bauswein:2011tp}. 
These relations could be used to tightly constrain
the NS EoS, especially if the EoS retains baryonic degrees of freedom
also at the highest densities
\cite{Bernuzzi:2015rla,
Breschi:2021xrx}. On the contrary, an observed violation
of these relations connecting the inspiral and postmerger features
could be an indication of a strong first order phase transition or of the
appearance of nucleonic resonances or hyperons at the postmerger densities
and temperatures 
\cite{Bauswein:2018bma, 
Radice:2016rys,Prakash:2021wpz}.
Crossover scenarios are discussed in 
Refs.~\cite{Fujimoto:2024ymt,Kapusta:2021ney,McLerran:2020rnw,Fujimoto:2022xhv}.
These GW measurements of NS matter 
are complementary \cite{Lovato:2022vgq} to the new laboratory probes of QCD at high densities 
that will take place 
at GSI/FAIR. 
In the Compressed Baryonic Matter experiment (CBM)~\cite{CBM:2016kpk,Galatyuk:2019lcf}, 
following previous High Acceptance DiElectron Spectrometer (HADES) measurements~\cite{HADES:2019auv},  
heavy ion collisions will produce QCD matter at similar densities to NS interiors with beam energies up to roughly 12 GeV per nucleon.
There are more protons in the heavy-ion collisions so this is one difference. 
The experiments will probe 
temperatures from about 70 MeV, the temperature expected in the excited NS merger process, to 200 MeV.
How might any phase transition work as a function of temperature at these densities? 
A crossover is expected from pure QCD in the early Universe at zero baryon densities (contrasting with the high densities in NSs). 
Information from kilonova observations will guide future nuclear structure experiments.
r-process nucleosynthesis occurs far from the valley of stability, inside
a region of the nuclear chart where no experimental information about
masses, reactions and decay rates are available. Thus, information from
kilonova observations will guide future nuclear structure experiments to
explore key points in the very neutron rich side of the valley of stability
at facilities like RIKEN, FRIB and GSI/FAIR.
The EoS of neutron rich matter including the role of $\Delta$ excitations and nucleosynthesis of heavy nuclei will be explored by the NuSTAR experiment 
in collisions up to 1.5 GeV/u \cite{Geissel:2003lcy,Rodriguez-Sanchez:2021zik}.

So far we have just one 
multi-messenger NS event.
The ET and CE are expected to detect about $10^4$ NS mergers up to redshift $z \sim 3$ with between 10s-100s of these per year measured in parallel with an electromagnetic signal 
leading to at least an order of magnitude improvement in precision on the NS EoS compared to LVK.
LISA with its lower frequency range will detect binary inspirals 
thousands to millions of years before the actual merger. 
One expects 
between on the order of 10s  and a few hundred NS (pre-)merger events over the mission lifetime.
There is also the possibility of axion DM condensates in NSs.
These would take longer to form than might be seen in heavy-ion experiments, e.g., at CBM.  
They could affect the NS EoS 
and would come with their own signature in GWs \cite{Zhang:2021mks,Kumamoto:2024wjd}.
Further insight would come from neutrino astrophysics.
Detecting a neutrino in temporal and directional coincidence with a
gravitational-wave event would be a major multimessenger
breakthrough, revealing details of compact-object mergers, relativistic
outflows, and dense matter \cite{IceCube:2026ads}. The IceCube neutrino experiment at the South Pole already participates in low-latency
GW follow-up. The IceCube upgrade will strengthen this role through improved pointing, reconstruction, and reduced systematics. These
advances enhance IceCube’s performance both as a rapid follow-up instrument and as an independent trigger for high-energy transients.
Strategies will allow the combination of the future extension IceCube-Gen2 \cite{IceCube-Gen2:2020qha} with next-generation GW facilities like the ET.

GWs can also shed light on the expansion rate of the Universe through the present day value of the Hubble parameter $H_0$ and DE.
Electromagnetic signal 
measurements come from  
the CMB through the Planck mission \cite{Planck:2018vyg}, 
the Atacama Cosmology Telescope \cite{ACT:2025tim}   
and the South Pole Telescope 
\cite{SPT-3G:2024qkd} 
at $z=1100$.
There are also measurements from 
Baryon Acoustic Oscillations, 
ripples in the distributions of galaxies left over from the big bang,  
which include the newest DESI, 
Dark Energy Spectroscopic Instrument,  data at $0.1 < z < 4.2$
\cite{DESI:2024mwx,DESI:2025zgx}. 
Recent time 
Cepheid measurements 
from $z \leq 0.01$ are outlined in \cite{Riess:2019qba}.
There is a tension between the value 
$H_0 = 67.4 \pm 0.5$ km/s/Mpc extracted from early Universe CMB measurements and the value 
$73.0 \pm 1.0$ km/s/Mpc extracted from late time Cepheid measurements.
For recent discussion see Refs.~\cite{Riess:2019qba,Efstathiou:2024dvn}.
In addition, recent DESI data hint at possible time dependent DE \cite{DESI:2024mwx,DESI:2025zgx}. 
Future measurements will come from the Vera Rubin Telescope and the Euclid experiment.
Merger events can give independent probes of $H_0$
and 
DE.
The GW170817 event gave a first GW measurement of $H_0$, though with large errors~\cite{LIGOScientific:2017adf}, 
$H_0 = 70^{+12}_{-8}$ km/s/Mpc.
This "bright siren" allows a simultaneous measurement of the luminosity distance (through GWs) and the redshift 
(through electromagnetic emission), and thus a direct measurement of the expansion rate of the Universe.
Besides neutron stars, mergers of supermassive black holes surrounded by gas clouds are also expected to produce an electromagnetic counterpart. 
These measurements plus 
BH-NS mergers in future will become more sensitive to 
DE with mergers away from redshifts $z \ll 1$.
LISA expects to achieve a precision of sub percent level for $H_0$
plus better than 10\% on the DE EoS, thus providing a fully independent measurement which may help resolve the Hubble tension and DE time (in-)dependence  
from measurements in the redshift range 
$0.01 \lesssim z \lesssim 7$ \cite{Caprini:2025mfr}.
Assuming that systematics including calibration are taken into account, 
the ET and CE should each give a sub-percent accurate measurement of $H_0$ plus much improved precision on the DE EoS.
Time dependent DE might imply a time dependent vacuum expectation value of 
some new scalar field \cite{Peebles:1987ek,Peebles:2002gy,Wetterich:1987fm,Wetterich:1994bg} 
or the vacuum relaxing away from an initial phase transition that might have produced the SM \cite{Bass:2023ece}.
With forthcoming precision measurements, there is the 
need to bring 
the GW and electromagnetic communities together to be able to do synthetic analyses. This is essential especially to do population studies for e.g. the NS EoS and $H_0$.  
The potential for electromagnetic-GW population studies is discussed in 
Refs.~\cite{Feeney:2018mkj,Feeney:2020kxk,Sarin:2023tgv}.
The power of these probes (both precision and accuracy) will be substantially impacted by the lack of redshifts.

\section{Stochastic gravitational wave backgrounds}

Beyond these merger event sources there are also SGWBs, e.g., as suggested in the few standard deviations 
effect observed with pulsar timing arrays. 
In future a few orders of magnitude sensitivity improvement at the nHz frequency range probed by pulsars is expected with the Square Kilometer Array \cite{Weltman:2018zrl}.
The LISA frequency band will be very interesting as a test for new physics at the TeV scale. 
SGWBs arise from the superposition of GWs
with different wave-vectors, frequencies and phases. 
SGWBs will be generated through first order phase transitions in the early Universe where bubbles of vacuum form and collide with a large release of energy in GWs \cite{Witten:1984rs,Hogan:1986dsh}. 
In contrast, 
crossover transitions give no SGWB \cite{Domcke:2024soc}.
SGWBs can also be generated by astrophysical sources like SMBHs at the centres of galaxies as unresolved sources which contribute to the signal measured in the detector but coming from outside the volume for individual detection. 
The effect seen in pulsar timing arrays is believed to come from SMBHs. 
Cosmological SGWBs will behave like the CMB appearing to the observer as coming from a spherical surface with radius the cosmic time of the emission event.
SGWBs from a first order phase transition 
in the early Universe will be seen peaked at a frequency 
\begin{equation}
    f_{\rm GW} = f_* \frac{a_*}{a_0} 
\approx
{\cal O}(1) \times 10^{-5} \frac{1}{H_{*} R_{\rm coll}} 
\frac{T_{*}}{100 {\rm GeV}} \ {\rm Hz} .
\end{equation}
Here $T_*$ and $H_*$ are the temperature and Hubble parameter at the end of the phase transition,  $R_{\rm coll}$ is the size of the bubbles at collision and the ${\cal O}(10^{-5})$ factor depends on the details of the source.
The frequency $f_{\rm GW}$ is obtained after gravitational redshift from the emission frequency $f_* = 1/R_{\rm coll} \geq H_*$ with 
$a_*$ and $a_0$ the Universe scale factors at the times of GW emission and detection, respectively.
The QCD and TeV scales correspond to frequencies in the nHz and LISA ranges respectively. 
A large energy scale $\sim 10^{16}$ GeV typical of inflation, emergence and unification models corresponds to frequencies around a GHz.

\begin{figure}[t!]  
\centerline
{\includegraphics[width=0.50\textwidth]
{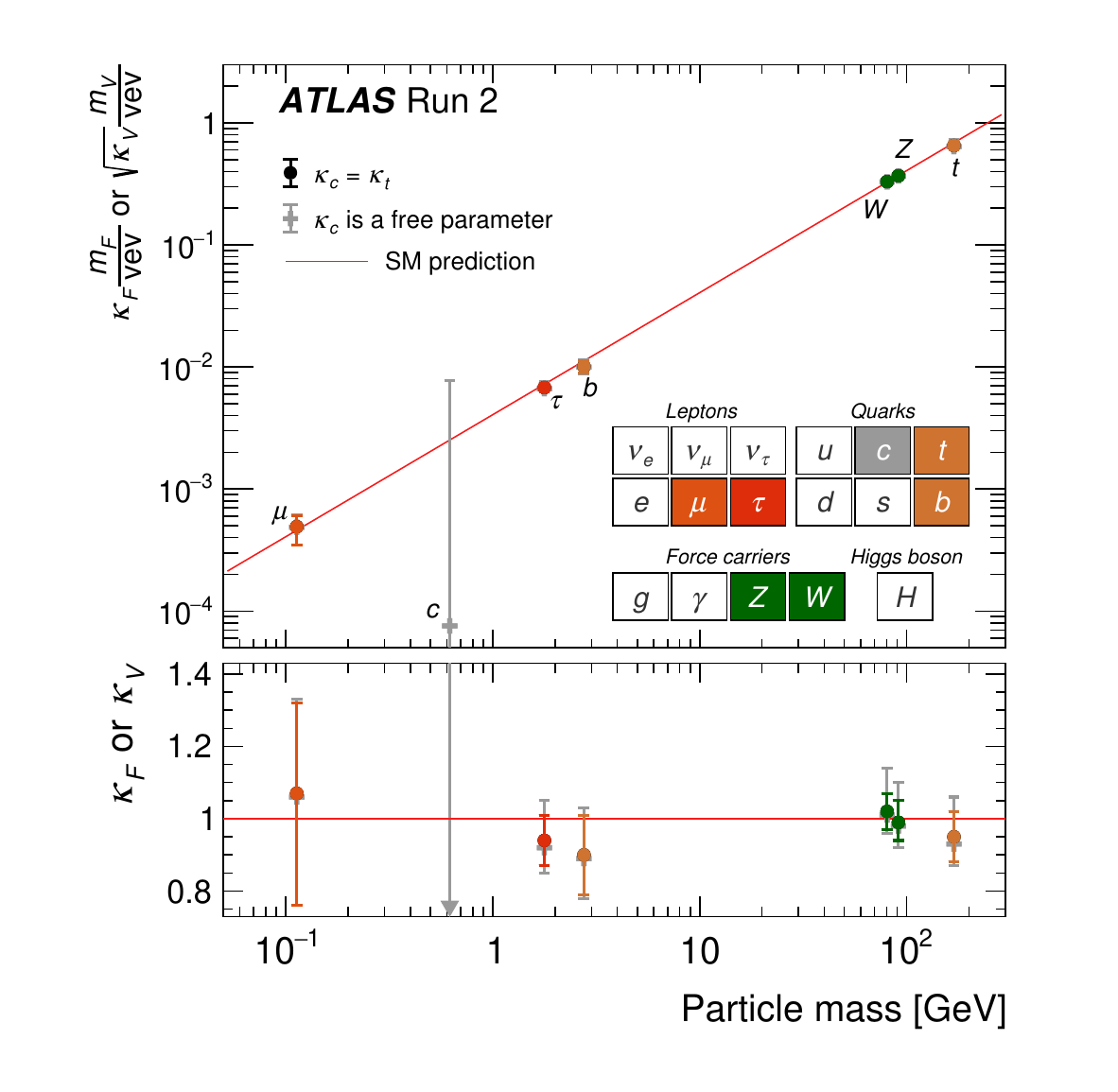} 
\hfill
\includegraphics[width=0.48\textwidth]
{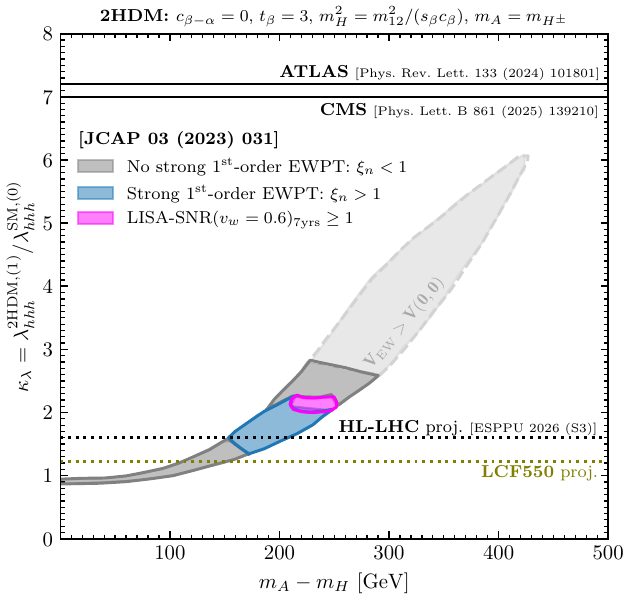}}
\caption{Left: Higgs boson couplings to different particles and masses measured at the LHC. The results are expressed in terms of "coupling modifiers" which describe the level of deviation from SM expectations \cite{ATLAS:2022vkf}.
Right: 
Parameter space of an extended 2-Higgs doublet model 
leading to a first-order 
phase transition in the LISA frequency range 
with a value of $\lambda$ that is larger than the SM value by a factor about two
\cite{Biekotter:2022kgf,LinearColliderVision:2025hlt}.
Present LHC constraints on $\lambda$ are shown as horizontal lines with results from ATLAS~\cite{ATLAS:2024ish}
and CMS~\cite{CMS:2024awa}.
Here $h$, $H$ and $A$ are the lightest, most SM-like, and heaviest neutral scalar bosons in the model; $A$ is an extra pseudoscalar boson which is taken to be degenerate with two extra charged Higgs states in the model.
The Higgs self-coupling $\lambda$ awaits precision measurement and is important both for vacuum stability and for determining any possible TeV scale SGWB.}
\label{fig:SM}
\end{figure}

The SM is commonly expected to give a crossover at the electroweak scale \cite{DOnofrio:2015gop}
implying that any SGWB detection here would be a signal for new physics \cite{Weir:2017wfa}. 
For example, one can get a first order transition in models with an extended scalar sector.
The SM is so far working exceptionally well in present LHC data and in measurements of rare decays.
The masses of the SM particles are determined by their couplings to the scalar Brout-Englert-Higgs, BEH, field, viz.
\begin{equation}
m_{\rm W}^2 = \frac{1}{4} g^2 v^2 , 
\ \ \ 
m_{\rm Z}^2 = \frac{1}{4} (g^2 + g'^2 )v^2 , 
\ \ \ 
m_{f} = y_{f} \frac{v}{\sqrt{2}} ,
\ \ \
m_{h}^2 = 2 \lambda v^2  .
\ \ \ \ \ \ \ \ \ \
\label{eqno3} 
\end{equation}
Here W and Z are the weak interaction gauge bosons, $f$ are the charged fermions with Yukawa couplings $y_f$ and $h$ is the Higgs boson; 
$g$ and $g'$ are the SU(2) and U(1) gauge couplings, 
$\lambda$ is the Higgs self coupling and $v$ is the Higgs vacuum expectation value.
The BEH potential is
\begin{equation}
    V(\Phi) = - \mu^2 \Phi^\dagger \Phi + \lambda (\Phi^\dagger \Phi)^2
\end{equation}
with $m_{h}^2 = 2 \mu^2$ and $\Phi$ the scalar Higgs doublet field of the SM.
The SM relations in Eq.~(\ref{eqno3})
for the 
top and bottom quarks, W and Z gauge bosons and $\tau$ and $\mu$ charged leptons 
are all working well to within about 10\% in present LHC data up to the end of Run 2 with 140 fb$^{-1}$
, see Fig.~\ref{fig:SM}.
These results follow from the success of the LHC with collaboration of machine, experiments, GRID computing and precision theoretical calculations.
With the HL-LHC upgrade, 
a factor of 20 more statistics is expected to come from the LHC. 
One of the prime goals is to determine the shape of the Higgs potential that is realised in nature and to understand the underlying physics giving rise to the Higgs potential. The trilinear Higgs self coupling $\lambda$ and the quartic Higgs self coupling (which in the SM is also given in terms of $\lambda$) are crucial ingredients for this. In extensions of the SM all scalar fields present in the model contribute to the Higgs potential, so that the shape of the Higgs potential of an extended Higgs sector may be vastly different from the SM case. Currently only weak experimental bounds exist on $\lambda$, allowing deviations from the SM value of several hundred percent~\cite{ATLAS:2025hhd,CMS:2024awa,ATLAS:2024ish}.
While the present experimental constraints on the quartic Higgs self-coupling are weaker than the theoretical bounds from perturbative unitarity, significant improvements are expected to be possible at the HL-LHC~\cite{Stylianou:2023tgg}.
The HL-LHC should reduce the uncertainty to about 28\% with 3 ab$^{-1}$ = 3000 fb$^{-1}$ of integrated luminosity~\cite{CMS:2025hfp}; 
5\% uncertainty could be possible with a future 100 TeV proton-proton collider~\cite{Jakobs:2023fxh}. 
A positive value of $\lambda$ is necessary for the stability of the Higgs vacuum.
The coupling $\lambda$ decreases under renormalisation group evolution with increasing energy. In the pure SM without coupling to undiscovered new particles, the vacuum sits close to the border of stable and metastable with $\lambda$ crossing zero not below about $10^{10}$ GeV \cite{Bednyakov:2015sca,Jegerlehner:2013cta,Degrassi:2012ry}.
The SM vacuum is within a 
few standard deviations of remaining stable up to the Planck scale 
\cite{Bednyakov:2015sca}.
Models with extra Higgs states, perhaps connected to baryogenesis, 
can predict a significantly different value of the self coupling $\lambda$ of the SM-like Higgs boson~\cite{Bahl:2022jnx} and, as explained above, have a more complicated vacuum structure.
In particular, an extended Higgs sector with a value of $\lambda$ that is bigger by a factor of about two can yield a first order phase transition with 
SGWB in the LISA frequency range \cite{Biekotter:2022kgf}. 
This is illustrated in the right panel of Fig.~\ref{fig:SM}.
If it is present, 
observing a SGWB with LISA will be an important experimental challenge. 
Even with guiding input from particle physics it will be difficult to predict the exact signal to look for.
SGWB reconstruction is highly non-unique. 
Collider observables and GW observables probe different effective parameters and null results at LISA will not strongly constrain many scenarios for physics beyond the SM.
SGWBs will appear in the detector like a "noise" contribution which must be disentangled from possible instrument noise. 
SGWBs are typically isotropic, unpolarised and Gaussian. 
With ground based interferometers one looks for cross-correlations between the detectors where instrument noise is expected to be largely uncorrelated. 
In pulsar timing arrays, time delays from different pulsars are correlated to look for the SGWB.
With LISA alone, there is the issue of separating the SGWB from instrument noise with just the one instrument. 
Ideas are discussed in Refs.~\cite{Muratore:2021uqj,Muratore:2022nbh,Muratore:2023gxh,Hartwig:2023pft,Alvey:2024uoc,Pozzoli:2023lgz,Caprini:2024hue}. 
Extra cross instrument calibration could come from the proposed 
Chinese GW detectors in space ~\cite{Gong:2021gvw}, 
namely
TianQin \cite{TianQin:2015yph,Luo:2025sos} and Tajii \cite{Ruan:2018tsw} 
which aim to measure GWs with frequencies between $10^{-4}$ and 1 Hz 
and hope to launch in about 2035.
In addition, 
the proposed Japanese mission DECIGO  \cite{Kawamura:2020pcg} in space aims at filling the frequency gap between 0.1 to 10 Hz.
If any SGWB can be detected in the experiments, 
the HL-LHC and 
LISA plus the Chinese missions will provide complementary windows on the few TeV region 
running in parallel in the next decade.
One should also be on the watch for evidence of any first order phase transition related to DM or DE due to new physics waiting to be discovered. 
At frequencies above LISA, 
the ET and CE 
expect to deliver a gain of four orders of magnitude increase in sensitivity to the energy density of SGWBs relative to LVK. 
Multi-frequency observations of a SGWB will also bring extra confidence as well as measuring the spectral characteristics to better precision.
Present LVK constraints are given in Ref.~\cite{LIGOScientific:2025kry}.
SGWBs of cosmological origin are probes of
the early Universe, and potential sources include first-order phase
transitions as well as topological defects. 
IceCube can detect such
topological defects directly as relics of the Big Bang \cite{Ackermann:2023gmd}. One example is
magnetic monopoles. These relics are much heavier than SM
particles and hence produce less bremsstrahlung or even pass the
detector at non-relativistic velocities, which are distinct features of
the detected signals.

\section{GWs as probes of physics at the highest scales}

HFGWs can probe 
energy scales much above collider energies if they can be detected.
There is vigorous R\&D aimed at  measurements with GW frequencies greater than about a GHz \cite{Aggarwal:2020olq,Berlin:2021txa,Franciolini:2022htd}.
Here there are no known astrophysical processes which might enter.
One would be looking for evidence of new phase transitions in the early Universe, PBH mergers or some unknown new physics exotica.
There are good reasons to believe that {\it something} interesting happens at very large energy scales. 
One is electroweak vacuum stability and the Higgs self-coupling perhaps 
crossing zero above about $10^{10}$ GeV. 
Here, the calculations are very sensitive to the details of perturbative SM radiative corrections and to the exact values of SM parameters. 
One would like (much) more precise values of the top and Higgs mass as well as the QCD coupling from future collider experiments as the main input parameters.
The experimental precision that can be achieved at a future FCC-ee are 7 MeV on the top quark mass 
(however, there is also theory uncertainty at the present level of 30 MeV), 4 MeV on the Higgs mass and 0.1\% on the QCD coupling $\alpha_s$~\cite{deBlas:2944678}. 
The criticality of the Higgs vacuum might be evidence for some type of topological phase transition in the ultraviolet with the SM and its gauge symmetries emergent below the energy scale of the phase transition.
Tiny neutrino masses  suggested by neutrino oscillation experiments 
are usually 
interpreted via the Weinberg operator or within see-saw models in terms of a large scale about $10^{16}$ GeV \cite{Weinberg:1979sa,Altarelli:2013tya}, 
which includes the range where $\lambda$ might cross zero.
While suggestive, 
neutrino oscillation experiments measure only neutrino mass differences.
They do not tell us the absolute mass scale, 
whether neutrinos are
Majorana (meaning that they would be their own antiparticles) or Dirac, 
nor provide a direct measurement of how heavy the large mass scale really is.
Direct probes of absolute neutrino masses are 
neutrinoless double $\beta$-decay experiments (which test for the Majorana nature of neutrinos) and the KATRIN experiment which 
provides a clean, model-independent anchor for the light neutrino mass scale. 
KATRIN studies the endpoint of the tritium $\beta$-decay
spectrum. 
The observable is the effective electron-neutrino mass \cite{KATRIN:2024cdt}  with the experiment aiming at sensitivity to 
$m_\nu \sim 0.3$ eV or less. 
Next generation 
neutrinoless double $\beta-$decay experiments should be sensitive to a Majorana mass parameter down to about 0.01 eV \cite{Baudis:2023pzu}.
Observations at these levels
would imply very high mass scales or alternative mass generation mechanisms. 
The DE scale describes the energy density of the vacuum perceived by gravitation in Einstein's equations of General Relativity. 
Theoretically, 
the tiny DE scale 0.002 eV 
extracted from astrophysics \cite{Planck:2018vyg,Bass:2023ece}
might enter at the same order as Majorana neutrino mass terms in a low energy expansion associated with an emergent SM characterised by this large  scale~\cite{Bass:2020egf,Bass:2021wxv,Bass:2023ece,Bjorken:2001pe,Jegerlehner:2013cta}. 
A similar large scale appears in single field inflation models
\cite{Baumann:2008bn}, in grand unified theories  
and also in axion models where it is connected with an axion decay constant parameter.
Might we find evidence for 
a first order phase transition deep in the ultraviolet?
The physics here is totally unknown from present experiments but could be probed with GWs. 
What will we find?

GW observables in the CMB can also probe the inflationary era.
The homogeneity, isotropy and flatness of the observed Universe are commonly interpreted as requiring an initial period of accelerating expansion called inflation where the Universe might have grown by a factor of $10^{26}$ in the first $10^{-33}$ seconds. 
The tensor-to-scalar ratio $r_{\rm ts}$ describes the ratio of the amplitude of primordial gravitational waves (tensor perturbations generated by GWs) to the amplitude of primordial density fluctuations (scalar perturbations) in the inflationary period of the very early Universe. 
It can be extracted from polarisation observables~\cite{Komatsu:2022nvu} -- so called B modes -- in the CMB.
In single scalar inflaton models the value of $r_{\rm ts}$ is connected to 
the scale of inflation via 
$V = (r_{\rm ts}/0.01)^{1/4} \times 10^{16}$ GeV \cite{Lyth:1996im,Baumann:2008bn}.
Presently $r_{\rm ts} < 0.032$ at 95\% confidence level  \cite{Tristram:2021tvh}.
The Atacama Cosmology Telescope and the Simons Observatory aim to reduce the error to $3 \times 10^{-3}$ with deep high resolution observations - see Ref.\cite{SimonsObservatory:2024gol}.
Further accuracy should come from the 
Japanese LiteBIRD satelite with full sky coverage \cite{LiteBIRD:2022cnt,LiteBIRD:2023iei} in the 2030s with the error reduced to $10^{-3}$.
If a signal for $r_{\rm ts}$ can be detected, it will give very interesting new information about the earliest times in the Universe.
The dynamics underlying inflation is an open theoretical puzzle.
In general, precision studies of CMB observables are an important tool to constrain different scenarios, 
namely to discriminate between different models of inflation and using the 
reheating temperature 
(the temperature at the end of the transition where the energy stored in inflation was converted to relativistic particles) 
to constrain particle physics scenarios including dark matter scenarios that are sensitive to it \cite{Ghoshal:2025ejg}.

GW signals from inflation might also be observable in next generation laser interferometers.  
Mergers and SGWBs from first order phase transitions come with peaked spectra. 
Broader spectra extended over several orders of magnitude can be induced in some inflation models.  
Inflation stretches the SGWB of primordial
GWs to cosmological distances. 
While the simplest single field slow-roll model of inflation 
gives too weak a signal to
be observed in 
experiments, 
more complicated scenarios might give a measurable signal. 
If there are additional active degrees besides just the inflaton, 
possible new symmetry patterns 
and/or if the scalar inflationary perturbations are sufficiently large to generate an interesting abundance of PBHs, 
then GW signals \cite{Maggiore:1999vm} can be generated  
which may be accessible to LISA and the ET and CE \cite{Bartolo:2016ami,Ricciardone:2016ddg,Guzzetti:2016mkm,Caprini:2018mtu,Maggiore:1999vm,Domcke:2024soc}.

\section{Future science perspectives and detector technologies}

Besides key physics observables, 
there are also important synergies between particle physics and GW detector technologies plus computing infrastructure investments waiting to be developed.
These include with the ET and CE 
plus new developments with atom interferometry and HFGWs detection.
In Europe the final ET configuration is under evaluation with the options being either a single 10 km length arms 
triangular site 
comprising three nested interferometers
or two 15 km L-shaped configurations in far separated regions. 
The science return for each configuration is 
discussed in Refs. \cite{Branchesi:2023mws,Maggiore:2024cwf,Abac:2025saz}.
The interferometers would be 200 - 300 m deep underground. 
Several sites have been proposed, 
with the present candidates being Sardinia, Saxony  and
the Euroregion Rhine-Meuse 
near the
Netherlands-Belgium-German border. 
The CE involves 2 L-shaped interferometers on the surface with lengths 20 km and 40 km similar to LIGO  
where proposed sites have still yet to be determined.
Sensitivity studies are reported in \cite{Hild:2010id}.
The ET will involve a two interferometer configuration in each arm
targeting high frequency (HF) GWs between 30 Hz and 2 kHz and low frequency (LF) GWs between about 2 Hz and 30 Hz.
The ET-HF will use high laser power and operate at room temperature, whereas the ET-LF will use low laser power and operate at cryogenic temperatures to suppress thermal noise.
Initially, 
the CE is planned to be 
high-power and non cryogenic. 
Cryogenic detectors might be introduced later with a stage-2 detector pioneered by the LIGO Voyager project  \cite{Reitze:2019dyk,Hall:2020dps}.
Synergies between the ET and CERN programmes 
are being developed, especially in vacuum and cryogenic technologies. 
On the more methodological side is the leveraging of artificial intelligence (AI) techniques with future collider data analysis and to better model complex astrophysics and/or speed up analyses so that we can extract fundamental physics more robustly with fewer approximations. 
The GW field has been pioneering in simulation-based inference \cite{Bhardwaj:2023xph,Alvey:2023npw}. 
The symbiotic co-evolution of physics-based simulation and AI emulation methods is essential for the future of the field with related 
implications for computing infrastructure investments in Europe.

In parallel to laser interferometers, 
there are ideas to use cold atom interferometers for GW and ultralight DM detection with a vigorous R\&D programme \cite{Badurina:2021rgt} and extending the use of quantum sensors in particle physics \cite{Bass:2023hoi}. 
The interferometers use Sr atomic clocks at temperatures of a few $\mu$K. 
These are accurate to one second in about 30 billion years. 
A passing GW will change the light travel time for one clock. Alternatively, a DM cloud can change the atom frequency with corresponding change in the atomic energy levels. 
R\&D plans involve starting with atom interferometers about 10m in length. 
These should be followed by few hundred meters length devices \cite{TVLBAIProto:2025wyn}, 
with one possibility involving the use of an LHC access shaft at CERN \cite{Baynham:2025pzm}, 
leading eventually to few thousand km separated satellites in space \cite{Alonso:2022oot}.
The aim is to have 100 m to km long baseline atom interferometers in the 2030s and a space mission in 2045+.
A long baseline atom interferometer in space 
has the potential to bridge the frequency gap between LVK and LISA 
if challenges 
with noise reduction can be overcome \cite{Badurina:2021rgt}.

High-frequency gravitational-wave (HFGW) detectors are an active area of research and development \cite{Aggarwal:2020olq,Aggarwal:2025noe,Schmieden:2023fzn}.
In electromagnetic cavities immersed in strong magnetic fields, HFGW signals can be probed through two complementary mechanisms. At lower frequencies, typically below the GeV scale, gravitational waves can mechanically couple to the cavity by deforming its boundaries, thereby inducing coupling between electromagnetic eigenmodes. At higher frequencies, above the GeV scale, detection can instead proceed via direct gravitational-wave photon conversion through the inverse Gertsenshtein effect, in which the gravitational wave converts directly into a photon.
There are also synergies with axion searches 
\cite{Domcke:2022rgu,Capdevilla:2025omb,Kim:2025izt}.
Possibilities using the GEO600 facility in Germany are discussed in 
Ref.~\cite{Jungkind:2025oqm}.
At the present early state of R\&D one needs a few orders of magnitude improvement in precision to reach the needed sensitivity. 
Similar level challenges appeared in the early days of laser interferometers before the present LVK 
experiments, which will soon upgrade to precision GW science with the ET and CE.

In conclusion, 
the interface of particle physics and gravitational waves with next generation experiments involves a 
rich science programme.
Physics synergies range from the QCD scale to be studied at GSI/FAIR through the TeV scale at the HL-LHC and future high energy colliders to inflation era dynamics and the particle physics of the early Universe.
Key GW observables involve BHs, NSs and possible SGWBs from first order phase transitions in the early Universe.
When further combined with cosmology observations, we can look forward to a new level of precision science and exciting new discoveries. 
As we move beyond initial discovery to precision science in each of TeV scale particle physics, QCD dense matter studies, GW measurements and cosmology observations, there will be an increasing need for new theoretical developments as well as new experiments.
In the era in which we hope to do fundamental physics with gravitational-wave probes, accuracy and robustness will be essential.
In particular,  some of the simplifying working assumptions that have so far been useful will need to be replaced by more realistic and systematically tested models for the interplay between black holes, galaxies, and dark matter.
Black hole population synthesis, the assumptions about black hole-galaxy coevolution, and waveform modelling will all become really crucial for interpretation and requires significant improvement. 
Likewise, pushing the precision frontier in particle physics over the broadest possible energy range will require improved accuracy  calculations. 
It may also require new principles and deeper conceptual understanding, especially if the Standard Model continues to hold beyond the regime where many theoretical ideas had anticipated the appearance of new physics, including at the interface between particle physics and cosmology.
This Perspectives article is an invitation for the next generation of excellent young scientists to join in
the quest to push these frontiers!

\section*{Acknowledgments}

This paper evolved from discussions at the 833.WE-Heraeus-Seminar in Kitzbühel, Austria. We thank the Wilhelm and Else Heraeus Foundation for support.

\bibliography{sample}

\begin{thebibliography}{100}
\urlstyle{rm}
\expandafter\ifx\csname url\endcsname\relax
  \def\url#1{\texttt{#1}}\fi
\expandafter\ifx\csname urlprefix\endcsname\relax\def\urlprefix{URL }\fi
\expandafter\ifx\csname doiprefix\endcsname\relax\def\doiprefix{DOI: }\fi
\providecommand{\bibinfo}[2]{#2}
\providecommand{\eprint}[2][]{\url{#2}}

\bibitem{ATLAS:2012yve}
\bibinfo{author}{Aad, G.} \emph{et~al.}
\newblock \bibinfo{journal}{\bibinfo{title}{{Observation of a new particle in the search for the Standard Model Higgs boson with the ATLAS detector at the LHC}}}.
\newblock {\emph{\JournalTitle{Phys. Lett. B}}} \textbf{\bibinfo{volume}{716}}, \bibinfo{pages}{1--29} (\bibinfo{year}{2012}).

\bibitem{CMS:2012qbp}
\bibinfo{author}{Chatrchyan, S.} \emph{et~al.}
\newblock \bibinfo{journal}{\bibinfo{title}{{Observation of a New Boson at a Mass of 125 GeV with the CMS Experiment at the LHC}}}.
\newblock {\emph{\JournalTitle{Phys. Lett. B}}} \textbf{\bibinfo{volume}{716}}, \bibinfo{pages}{30--61} (\bibinfo{year}{2012}).

\bibitem{Bass:2021acr}
\bibinfo{author}{Bass, S.~D.}, \bibinfo{author}{De~Roeck, A.} \& \bibinfo{author}{Kado, M.}
\newblock \bibinfo{journal}{\bibinfo{title}{{The Higgs boson implications and prospects for future discoveries}}}.
\newblock {\emph{\JournalTitle{Nature Rev. Phys.}}} \textbf{\bibinfo{volume}{3}}, \bibinfo{pages}{608--624} (\bibinfo{year}{2021}).

\bibitem{Jakobs:2023fxh}
\bibinfo{author}{Jakobs, K.} \& \bibinfo{author}{Zanderighi, G.}
\newblock \bibinfo{journal}{\bibinfo{title}{{The profile of the Higgs boson: status and prospects}}}.
\newblock {\emph{\JournalTitle{Phil. Trans. Roy. Soc. Lond. A}}} \textbf{\bibinfo{volume}{382}}, \bibinfo{pages}{20230087} (\bibinfo{year}{2023}).

\bibitem{deBlas:2944678}
\bibinfo{author}{de~Blas, J.} \emph{et~al.}
\newblock \bibinfo{journal}{\bibinfo{title}{{Physics Briefing Book: Input for the 2026 update of the European Strategy for Particle Physics}}}.
\newblock \doiprefix\url{10.17181/CERN.35CH.2O2P} (\bibinfo{year}{2025}).
\newblock \eprint{2511.03883}.

\bibitem{LIGOScientific:2025slb}
\bibinfo{author}{Abac, A.~G.} \emph{et~al.}
\newblock \bibinfo{journal}{\bibinfo{title}{{GWTC-4.0: Updating the Gravitational-Wave Transient Catalog with Observations from the First Part of the Fourth LIGO-Virgo-KAGRA Observing Run}}}.
\newblock (\bibinfo{year}{2025}).
\newblock \eprint{2508.18082}.

\bibitem{BHweb}
\bibinfo{note}{{https://gracedb.ligo.org/superevents/public/O4/}}.

\bibitem{LIGOScientific:2017vwq}
\bibinfo{author}{Abbott, B.~P.} \emph{et~al.}
\newblock \bibinfo{journal}{\bibinfo{title}{{GW170817: Observation of Gravitational Waves from a Binary Neutron Star Inspiral}}}.
\newblock {\emph{\JournalTitle{Phys. Rev. Lett.}}} \textbf{\bibinfo{volume}{119}}, \bibinfo{pages}{161101} (\bibinfo{year}{2017}).

\bibitem{LIGOScientific:2017zic}
\bibinfo{author}{Abbott, B.~P.} \emph{et~al.}
\newblock \bibinfo{journal}{\bibinfo{title}{{Gravitational Waves and Gamma-rays from a Binary Neutron Star Merger: GW170817 and GRB 170817A}}}.
\newblock {\emph{\JournalTitle{Astrophys. J. Lett.}}} \textbf{\bibinfo{volume}{848}}, \bibinfo{pages}{L13} (\bibinfo{year}{2017}).

\bibitem{LIGOScientific:2014qfs}
\bibinfo{author}{Aasi, J.} \emph{et~al.}
\newblock \bibinfo{journal}{\bibinfo{title}{{Characterization of the LIGO detectors during their sixth science run}}}.
\newblock {\emph{\JournalTitle{Class. Quant. Grav.}}} \textbf{\bibinfo{volume}{32}}, \bibinfo{pages}{115012} (\bibinfo{year}{2015}).

\bibitem{VIRGO:2014yos}
\bibinfo{author}{Acernese, F.} \emph{et~al.}
\newblock \bibinfo{journal}{\bibinfo{title}{{Advanced Virgo: a second-generation interferometric gravitational wave detector}}}.
\newblock {\emph{\JournalTitle{Class. Quant. Grav.}}} \textbf{\bibinfo{volume}{32}}, \bibinfo{pages}{024001} (\bibinfo{year}{2015}).

\bibitem{KAGRA:2018plz}
\bibinfo{author}{Akutsu, T.} \emph{et~al.}
\newblock \bibinfo{journal}{\bibinfo{title}{{KAGRA: 2.5 Generation Interferometric Gravitational Wave Detector}}}.
\newblock {\emph{\JournalTitle{Nature Astron.}}} \textbf{\bibinfo{volume}{3}}, \bibinfo{pages}{35--40} (\bibinfo{year}{2019}).

\bibitem{Heurs:2018wsu}
\bibinfo{author}{Heurs, M.}
\newblock \bibinfo{journal}{\bibinfo{title}{{Gravitational wave detection using laser interferometry beyond the standard quantum limit}}}.
\newblock {\emph{\JournalTitle{Phil. Trans. Roy. Soc. Lond. A}}} \textbf{\bibinfo{volume}{376}}, \bibinfo{pages}{20170289} (\bibinfo{year}{2018}).

\bibitem{NANOGrav:2023gor}
\bibinfo{author}{Agazie, G.} \emph{et~al.}
\newblock \bibinfo{journal}{\bibinfo{title}{{The NANOGrav 15 yr Data Set: Evidence for a Gravitational-wave Background}}}.
\newblock {\emph{\JournalTitle{Astrophys. J. Lett.}}} \textbf{\bibinfo{volume}{951}}, \bibinfo{pages}{L8} (\bibinfo{year}{2023}).

\bibitem{EPTA:2023fyk}
\bibinfo{author}{Antoniadis, J.} \emph{et~al.}
\newblock \bibinfo{journal}{\bibinfo{title}{{The second data release from the European Pulsar Timing Array - III. Search for gravitational wave signals}}}.
\newblock {\emph{\JournalTitle{Astron. Astrophys.}}} \textbf{\bibinfo{volume}{678}}, \bibinfo{pages}{A50} (\bibinfo{year}{2023}).

\bibitem{Reardon:2023gzh}
\bibinfo{author}{Reardon, D.~J.} \emph{et~al.}
\newblock \bibinfo{journal}{\bibinfo{title}{{Search for an Isotropic Gravitational-wave Background with the Parkes Pulsar Timing Array}}}.
\newblock {\emph{\JournalTitle{Astrophys. J. Lett.}}} \textbf{\bibinfo{volume}{951}}, \bibinfo{pages}{L6} (\bibinfo{year}{2023}).

\bibitem{Xu:2023wog}
\bibinfo{author}{Xu, H.} \emph{et~al.}
\newblock \bibinfo{journal}{\bibinfo{title}{{Searching for the Nano-Hertz Stochastic Gravitational Wave Background with the Chinese Pulsar Timing Array Data Release I}}}.
\newblock {\emph{\JournalTitle{Res. Astron. Astrophys.}}} \textbf{\bibinfo{volume}{23}}, \bibinfo{pages}{075024} (\bibinfo{year}{2023}).

\bibitem{Caprini:2024lxj}
\bibinfo{author}{Caprini, C.}
\newblock \bibinfo{journal}{\bibinfo{title}{{Strong evidence for the discovery of a gravitational wave background}}}.
\newblock {\emph{\JournalTitle{Nature Rev. Phys.}}} \textbf{\bibinfo{volume}{6}}, \bibinfo{pages}{291--293} (\bibinfo{year}{2024}).

\bibitem{Bailes:2021tot}
\bibinfo{author}{Bailes, M.} \emph{et~al.}
\newblock \bibinfo{journal}{\bibinfo{title}{{Gravitational-wave physics and astronomy in the 2020s and 2030s}}}.
\newblock {\emph{\JournalTitle{Nature Rev. Phys.}}} \textbf{\bibinfo{volume}{3}}, \bibinfo{pages}{344--366} (\bibinfo{year}{2021}).

\bibitem{LISA:2024hlh}
\bibinfo{author}{Colpi, M.} \emph{et~al.}
\newblock \bibinfo{journal}{\bibinfo{title}{{LISA Definition Study Report}}}.
\newblock (\bibinfo{year}{2024}).
\newblock \eprint{2402.07571}.

\bibitem{Caprini:2025mfr}
\bibinfo{author}{Caprini, C.} \emph{et~al.}
\newblock \bibinfo{journal}{\bibinfo{title}{{Science of the LISA mission: A Summary for the European Strategy for Particle Physics}}}.
\newblock (\bibinfo{year}{2025}).
\newblock \eprint{2507.05130}.

\bibitem{LISAPathfinder:2024ucp}
\bibinfo{author}{Armano, M.} \emph{et~al.}
\newblock \bibinfo{journal}{\bibinfo{title}{{In-depth analysis of LISA Pathfinder performance results: Time evolution, noise projection, physical models, and implications for LISA}}}.
\newblock {\emph{\JournalTitle{Phys. Rev. D}}} \textbf{\bibinfo{volume}{110}}, \bibinfo{pages}{042004} (\bibinfo{year}{2024}).

\bibitem{LISA:2022kgy}
\bibinfo{author}{Arun, K.~G.} \emph{et~al.}
\newblock \bibinfo{journal}{\bibinfo{title}{{New horizons for fundamental physics with LISA}}}.
\newblock {\emph{\JournalTitle{Living Rev. Rel.}}} \textbf{\bibinfo{volume}{25}}, \bibinfo{pages}{4} (\bibinfo{year}{2022}).

\bibitem{LISACosmologyWorkingGroup:2022jok}
\bibinfo{author}{Auclair, P.} \emph{et~al.}
\newblock \bibinfo{journal}{\bibinfo{title}{{Cosmology with the Laser Interferometer Space Antenna}}}.
\newblock {\emph{\JournalTitle{Living Rev. Rel.}}} \textbf{\bibinfo{volume}{26}}, \bibinfo{pages}{5} (\bibinfo{year}{2023}).

\bibitem{LISA:2022yao}
\bibinfo{author}{Seoane, P.~A.} \emph{et~al.}
\newblock \bibinfo{journal}{\bibinfo{title}{{Astrophysics with the Laser Interferometer Space Antenna}}}.
\newblock {\emph{\JournalTitle{Living Rev. Rel.}}} \textbf{\bibinfo{volume}{26}}, \bibinfo{pages}{2} (\bibinfo{year}{2023}).

\bibitem{Branchesi:2023mws}
\bibinfo{author}{Branchesi, M.} \emph{et~al.}
\newblock \bibinfo{journal}{\bibinfo{title}{{Science with the Einstein Telescope: a comparison of different designs}}}.
\newblock {\emph{\JournalTitle{JCAP}}} \textbf{\bibinfo{volume}{07}}, \bibinfo{pages}{068} (\bibinfo{year}{2023}).

\bibitem{Abac:2025saz}
\bibinfo{author}{Abac, A.} \emph{et~al.}
\newblock \bibinfo{journal}{\bibinfo{title}{{The Science of the Einstein Telescope}}}.
\newblock (\bibinfo{year}{2025}).
\newblock \eprint{2503.12263}.

\bibitem{ET:2019dnz}
\bibinfo{author}{Maggiore, M.} \emph{et~al.}
\newblock \bibinfo{journal}{\bibinfo{title}{{Science Case for the Einstein Telescope}}}.
\newblock {\emph{\JournalTitle{JCAP}}} \textbf{\bibinfo{volume}{03}}, \bibinfo{pages}{050} (\bibinfo{year}{2020}).

\bibitem{Evans:2021gyd}
\bibinfo{author}{Evans, M.} \emph{et~al.}
\newblock \bibinfo{journal}{\bibinfo{title}{{A Horizon Study for Cosmic Explorer: Science, Observatories, and Community}}}.
\newblock (\bibinfo{year}{2021}).
\newblock \eprint{2109.09882}.

\bibitem{Evans:2023euw}
\bibinfo{author}{Evans, M.} \emph{et~al.}
\newblock \bibinfo{journal}{\bibinfo{title}{{Cosmic Explorer: A Submission to the NSF MPSAC ngGW Subcommittee}}}.
\newblock (\bibinfo{year}{2023}).
\newblock \eprint{2306.13745}.

\bibitem{TVLBAISummary1}
\bibinfo{author}{Abend, S.} \emph{et~al.}
\newblock \bibinfo{journal}{\bibinfo{title}{Terrestrial very-long-baseline atom interferometry}}.
\newblock {\emph{\JournalTitle{AVS Quantum Science}}} \textbf{\bibinfo{volume}{6}}, \bibinfo{pages}{024701} (\bibinfo{year}{2023}).

\bibitem{TVLBAISummary2}
\bibinfo{author}{Abdalla, A.} \emph{et~al.}
\newblock \bibinfo{journal}{\bibinfo{title}{Terrestrial {Very}-{Long}-{Baseline} {Atom} {Interferometry}: summary of the second workshop}}.
\newblock {\emph{\JournalTitle{EPJ Quantum Technology}}} \textbf{\bibinfo{volume}{12}}, \bibinfo{pages}{42} (\bibinfo{year}{2025}).

\bibitem{schlippert2020matter}
\bibinfo{author}{Schlippert, D.} \emph{et~al.}
\newblock \bibinfo{title}{Matter-wave interferometry for inertial sensing and tests of fundamental physics}.
\newblock In \emph{\bibinfo{booktitle}{CPT AND LORENTZ SYMMETRY: Proceedings of the Eighth Meeting on CPT and Lorentz Symmetry}}, \bibinfo{pages}{37--40} (\bibinfo{organization}{World Scientific}, \bibinfo{year}{2020}).

\bibitem{Overstreet:2021hea}
\bibinfo{author}{Overstreet, C.}, \bibinfo{author}{Asenbaum, P.}, \bibinfo{author}{Curti, J.}, \bibinfo{author}{Kim, M.} \& \bibinfo{author}{Kasevich, M.~A.}
\newblock \bibinfo{journal}{\bibinfo{title}{{Observation of a gravitational Aharonov-Bohm effect}}}.
\newblock {\emph{\JournalTitle{Science}}} \textbf{\bibinfo{volume}{375}}, \bibinfo{pages}{abl7152} (\bibinfo{year}{2021}).

\bibitem{zhou2011development}
\bibinfo{author}{Zhou, L.} \emph{et~al.}
\newblock \bibinfo{journal}{\bibinfo{title}{Development of an atom gravimeter and status of the 10-meter atom interferometer for precision gravity measurement}}.
\newblock {\emph{\JournalTitle{General Relativity and Gravitation}}} \textbf{\bibinfo{volume}{43}}, \bibinfo{pages}{1931--1942} (\bibinfo{year}{2011}).

\bibitem{Bongs:2025rqe}
\bibinfo{author}{Bongs, K.} \emph{et~al.}
\newblock \bibinfo{journal}{\bibinfo{title}{{AION-10: Technical Design Report for a 10m Atom Interferometer in Oxford}}}.
\newblock  (\bibinfo{year}{2025}).
\newblock \eprint{2508.03491}.

\bibitem{MAGIS-100:2021etm}
\bibinfo{author}{Abe, M.} \emph{et~al.}
\newblock \bibinfo{journal}{\bibinfo{title}{{Matter-wave Atomic Gradiometer Interferometric Sensor (MAGIS-100)}}}.
\newblock {\emph{\JournalTitle{Quantum Sci. Technol.}}} \textbf{\bibinfo{volume}{6}}, \bibinfo{pages}{044003} (\bibinfo{year}{2021}).

\bibitem{Canuel:2017rrp}
\bibinfo{author}{Canuel, B.} \emph{et~al.}
\newblock \bibinfo{journal}{\bibinfo{title}{{Exploring gravity with the MIGA large scale atom interferometer}}}.
\newblock {\emph{\JournalTitle{Sci. Rep.}}} \textbf{\bibinfo{volume}{8}}, \bibinfo{pages}{14064} (\bibinfo{year}{2018}).

\bibitem{Baynham:2025pzm}
\bibinfo{author}{Baynham, C.} \emph{et~al.}
\newblock \bibinfo{journal}{\bibinfo{title}{{Letter of Intent: AICE -- 100m Atom Interferometer Experiment at CERN}}}.
\newblock (\bibinfo{year}{2025}).
\newblock \eprint{2509.11867}.

\bibitem{Aggarwal:2020olq}
\bibinfo{author}{Aggarwal, N.} \emph{et~al.}
\newblock \bibinfo{journal}{\bibinfo{title}{{Challenges and opportunities of gravitational-wave searches at MHz to GHz frequencies}}}.
\newblock {\emph{\JournalTitle{Living Rev. Rel.}}} \textbf{\bibinfo{volume}{24}}, \bibinfo{pages}{4} (\bibinfo{year}{2021}).

\bibitem{Berlin:2021txa}
\bibinfo{author}{Berlin, A.} \emph{et~al.}
\newblock \bibinfo{journal}{\bibinfo{title}{{Detecting high-frequency gravitational waves with microwave cavities}}}.
\newblock {\emph{\JournalTitle{Phys. Rev. D}}} \textbf{\bibinfo{volume}{105}}, \bibinfo{pages}{116011} (\bibinfo{year}{2022}).

\bibitem{Franciolini:2022htd}
\bibinfo{author}{Franciolini, G.}, \bibinfo{author}{Maharana, A.} \& \bibinfo{author}{Muia, F.}
\newblock \bibinfo{journal}{\bibinfo{title}{{Hunt for light primordial black hole dark matter with ultrahigh-frequency gravitational waves}}}.
\newblock {\emph{\JournalTitle{Phys. Rev. D}}} \textbf{\bibinfo{volume}{106}}, \bibinfo{pages}{103520} (\bibinfo{year}{2022}).

\bibitem{Badurina:2021rgt}
\bibinfo{author}{Badurina, L.} \emph{et~al.}
\newblock \bibinfo{journal}{\bibinfo{title}{{Prospective sensitivities of atom interferometers to gravitational waves and ultralight dark matter}}}.
\newblock {\emph{\JournalTitle{Phil. Trans. A. Math. Phys. Eng. Sci.}}} \textbf{\bibinfo{volume}{380}}, \bibinfo{pages}{20210060} (\bibinfo{year}{2021}).

\bibitem{Kawamura:2020pcg}
\bibinfo{author}{Kawamura, S.} \emph{et~al.}
\newblock \bibinfo{journal}{\bibinfo{title}{{Current status of space gravitational wave antenna DECIGO and B-DECIGO}}}.
\newblock {\emph{\JournalTitle{PTEP}}} \textbf{\bibinfo{volume}{2021}}, \bibinfo{pages}{05A105} (\bibinfo{year}{2021}).

\bibitem{LIGOScientific:2016aoc}
\bibinfo{author}{Abbott, B.~P.} \emph{et~al.}
\newblock \bibinfo{journal}{\bibinfo{title}{{Observation of Gravitational Waves from a Binary Black Hole Merger}}}.
\newblock {\emph{\JournalTitle{Phys. Rev. Lett.}}} \textbf{\bibinfo{volume}{116}}, \bibinfo{pages}{061102} (\bibinfo{year}{2016}).

\bibitem{LIGOScientific:2020iuh}
\bibinfo{author}{Abbott, R.} \emph{et~al.}
\newblock \bibinfo{journal}{\bibinfo{title}{{GW190521: A Binary Black Hole Merger with a Total Mass of $150 M_{\odot}$}}}.
\newblock {\emph{\JournalTitle{Phys. Rev. Lett.}}} \textbf{\bibinfo{volume}{125}}, \bibinfo{pages}{101102} (\bibinfo{year}{2020}).

\bibitem{LIGOScientific:2025rsn}
\bibinfo{author}{Abac, A.~G.} \emph{et~al.}
\newblock \bibinfo{journal}{\bibinfo{title}{{GW231123: a Binary Black Hole Merger with Total Mass 190-265 $M_{\odot}$}}}.
\newblock {\emph{\JournalTitle{Astrophys. J. Lett.}}} \textbf{\bibinfo{volume}{993}}, \bibinfo{pages}{L25} (\bibinfo{year}{2025}).

\bibitem{Genzel:2021lmp}
\bibinfo{author}{Genzel, R.}
\newblock \bibinfo{journal}{\bibinfo{title}{{A Forty Year Journey}}}.
\newblock {\emph{\JournalTitle{Rev. Mod. Phys.}}} \textbf{\bibinfo{volume}{94}}, \bibinfo{pages}{020501} (\bibinfo{year}{2022}).

\bibitem{Ferrarese:2002ct}
\bibinfo{author}{Ferrarese, L.}
\newblock \bibinfo{journal}{\bibinfo{title}{{Beyond the bulge: a fundamental relation between supermassive black holes and dark matter halos}}}.
\newblock {\emph{\JournalTitle{Astrophys. J.}}} \textbf{\bibinfo{volume}{578}}, \bibinfo{pages}{90--97} (\bibinfo{year}{2002}).

\bibitem{Ferrarese:2004qr}
\bibinfo{author}{Ferrarese, L.} \& \bibinfo{author}{Ford, H.}
\newblock \bibinfo{journal}{\bibinfo{title}{{Supermassive black holes in galactic nuclei: Past, present and future research}}}.
\newblock {\emph{\JournalTitle{Space Sci. Rev.}}} \textbf{\bibinfo{volume}{116}}, \bibinfo{pages}{523--624} (\bibinfo{year}{2005}).

\bibitem{Ferrarese:2006fd}
\bibinfo{author}{Ferrarese, L.} \emph{et~al.}
\newblock \bibinfo{journal}{\bibinfo{title}{{A fundamental relation between compact stellar nuclei, supermassive black holes, and their host galaxies}}}.
\newblock {\emph{\JournalTitle{Astrophys. J. Lett.}}} \textbf{\bibinfo{volume}{644}}, \bibinfo{pages}{L21--L24} (\bibinfo{year}{2006}).

\bibitem{Genzel:2024vou}
\bibinfo{author}{Genzel, R.}, \bibinfo{author}{Eisenhauer, F.} \& \bibinfo{author}{Gillessen, S.}
\newblock \bibinfo{journal}{\bibinfo{title}{{Experimental studies of black holes: status and future prospects}}}.
\newblock {\emph{\JournalTitle{Astron. Astrophys. Rev.}}} \textbf{\bibinfo{volume}{32}}, \bibinfo{pages}{3} (\bibinfo{year}{2024}).

\bibitem{2025ApJ...989L...7T}
\bibinfo{author}{{Taylor}, A.~J.} \emph{et~al.}
\newblock \bibinfo{journal}{\bibinfo{title}{{CAPERS-LRD-z9: A Gas-enshrouded Little Red Dot Hosting a Broad-line Active Galactic Nucleus at z = 9.288}}}.
\newblock {\emph{\JournalTitle{APJL}}} \textbf{\bibinfo{volume}{989}}, \bibinfo{pages}{L7} (\bibinfo{year}{2025}).

\bibitem{Maiolino:2025tih}
\bibinfo{author}{Maiolino, R.} \emph{et~al.}
\newblock \bibinfo{journal}{\bibinfo{title}{{A black hole in a near-pristine galaxy 700 million years after the Big Bang}}}.
\newblock  (\bibinfo{year}{2025}).
\newblock \eprint{2505.22567}.

\bibitem{Reynolds:2013rva}
\bibinfo{author}{Reynolds, C.~S.}
\newblock \bibinfo{journal}{\bibinfo{title}{{The Spin of Supermassive Black Holes}}}.
\newblock {\emph{\JournalTitle{Class. Quant. Grav.}}} \textbf{\bibinfo{volume}{30}}, \bibinfo{pages}{244004} (\bibinfo{year}{2013}).

\bibitem{McKernan:2023xio}
\bibinfo{author}{McKernan, B.} \& \bibinfo{author}{Ford, K. E.~S.}
\newblock \bibinfo{journal}{\bibinfo{title}{{Constraining the LVK AGN channel with black hole spins}}}.
\newblock {\emph{\JournalTitle{Mon. Not. Roy. Astron. Soc.}}} \textbf{\bibinfo{volume}{531}}, \bibinfo{pages}{3479--3485} (\bibinfo{year}{2024}).

\bibitem{Daly:2023axh}
\bibinfo{author}{Daly, R.~A.} \emph{et~al.}
\newblock \bibinfo{journal}{\bibinfo{title}{{New black hole spin values for Sagittarius A* obtained with the outflow method}}}.
\newblock {\emph{\JournalTitle{Mon. Not. Roy. Astron. Soc.}}} \textbf{\bibinfo{volume}{527}}, \bibinfo{pages}{428--436} (\bibinfo{year}{2023}).

\bibitem{Drew:2025euq}
\bibinfo{author}{Drew, M.}, \bibinfo{author}{Stanway, J.~S.}, \bibinfo{author}{Patterson, B.~A.}, \bibinfo{author}{Walton, T.~J.} \& \bibinfo{author}{Ward-Thompson, D.}
\newblock \bibinfo{journal}{\bibinfo{title}{{New Estimates of the Spin and Accretion Rate of the Black Hole M87*}}}.
\newblock {\emph{\JournalTitle{Astrophys. J. Lett.}}} \textbf{\bibinfo{volume}{984}}, \bibinfo{pages}{L31} (\bibinfo{year}{2025}).

\bibitem{Genzel:2017jgd}
\bibinfo{author}{Genzel, R.} \emph{et~al.}
\newblock \bibinfo{journal}{\bibinfo{title}{{Strongly baryon-dominated disk galaxies at the peak of galaxy formation ten billion years ago}}}.
\newblock {\emph{\JournalTitle{Nature}}} \textbf{\bibinfo{volume}{543}}, \bibinfo{pages}{397--401} (\bibinfo{year}{2017}).

\bibitem{Volonteri:2021sfo}
\bibinfo{author}{Volonteri, M.}, \bibinfo{author}{Habouzit, M.} \& \bibinfo{author}{Colpi, M.}
\newblock \bibinfo{journal}{\bibinfo{title}{{The origins of massive black holes}}}.
\newblock {\emph{\JournalTitle{Nature Rev. Phys.}}} \textbf{\bibinfo{volume}{3}}, \bibinfo{pages}{732--743} (\bibinfo{year}{2021}).

\bibitem{Baudis:PDG}
\bibinfo{author}{Baudis, L.} \& \bibinfo{author}{Profumom, S.}
\newblock \bibinfo{title}{Dark matter}.
\newblock In \bibinfo{editor}{Navas, S.} \emph{et~al.} (eds.) \emph{\bibinfo{booktitle}{{The Review of Particle Physics (2023), {\rm [Particle Data Group] (Chapter 27), Phys. Rev. D \textbf{110} (2024), 030001.}}}} (\bibinfo{year}{2024}).

\bibitem{Bertone:2004pz}
\bibinfo{author}{Bertone, G.}, \bibinfo{author}{Hooper, D.} \& \bibinfo{author}{Silk, J.}
\newblock \bibinfo{journal}{\bibinfo{title}{{Particle dark matter: Evidence, candidates and constraints}}}.
\newblock {\emph{\JournalTitle{Phys. Rept.}}} \textbf{\bibinfo{volume}{405}}, \bibinfo{pages}{279--390} (\bibinfo{year}{2005}).

\bibitem{Bertone:2018krk}
\bibinfo{author}{Bertone, G.} \& \bibinfo{author}{Tait, T. M.~P.}
\newblock \bibinfo{journal}{\bibinfo{title}{{A new era in the search for dark matter}}}.
\newblock {\emph{\JournalTitle{Nature}}} \textbf{\bibinfo{volume}{562}}, \bibinfo{pages}{51--56} (\bibinfo{year}{2018}).

\bibitem{DeRoeck:2024fjq}
\bibinfo{author}{De~Roeck, A.}
\newblock \bibinfo{journal}{\bibinfo{title}{{Dark matter searches at accelerators}}}.
\newblock {\emph{\JournalTitle{Nucl. Phys. B}}} \textbf{\bibinfo{volume}{1003}}, \bibinfo{pages}{116480} (\bibinfo{year}{2024}).

\bibitem{Baudis:2023pzu}
\bibinfo{author}{Baudis, L.}
\newblock \bibinfo{journal}{\bibinfo{title}{{Dual-phase xenon time projection chambers for~rare-event searches}}}.
\newblock {\emph{\JournalTitle{Phil. Trans. Roy. Soc. A}}} \textbf{\bibinfo{volume}{382}}, \bibinfo{pages}{20230083} (\bibinfo{year}{2023}).

\bibitem{Aprile:2024fkg}
\bibinfo{author}{Aprile, E.}
\newblock \bibinfo{journal}{\bibinfo{title}{{The XENON program for dark matter direct detection}}}.
\newblock {\emph{\JournalTitle{Nucl. Phys. B}}} \textbf{\bibinfo{volume}{1003}}, \bibinfo{pages}{116463} (\bibinfo{year}{2024}).

\bibitem{Dobrich:2025oso}
\bibinfo{author}{D{\"o}brich, B.} \& \bibinfo{author}{Irastorza, I.~G.}
\newblock \bibinfo{journal}{\bibinfo{title}{{Experiments to test the hypothesis for solar and dark matter axions}}}.
\newblock  (\bibinfo{year}{2025}).
\newblock \eprint{2507.06414}.

\bibitem{Goudzovski:2022vbt}
\bibinfo{author}{Goudzovski, E.} \emph{et~al.}
\newblock \bibinfo{journal}{\bibinfo{title}{{New physics searches at kaon and hyperon factories}}}.
\newblock {\emph{\JournalTitle{Rept. Prog. Phys.}}} \textbf{\bibinfo{volume}{86}}, \bibinfo{pages}{016201} (\bibinfo{year}{2023}).

\bibitem{NA62:2025upx}
\bibinfo{author}{Cortina~Gil, E.} \emph{et~al.}
\newblock \bibinfo{journal}{\bibinfo{title}{{Searches for hidden sectors using K$^{+}${\textrightarrow} {\ensuremath{\pi}}$^{+}$X decays}}}.
\newblock {\emph{\JournalTitle{JHEP}}} \textbf{\bibinfo{volume}{11}}, \bibinfo{pages}{143} (\bibinfo{year}{2025}).

\bibitem{LIGOScientific:2025ttj}
\bibinfo{author}{Abac, A.~G.} \emph{et~al.}
\newblock \bibinfo{journal}{\bibinfo{title}{{Direct multi-model dark-matter search with gravitational-wave interferometers using data from the first part of the fourth LIGO-Virgo-KAGRA observing run}}}.
\newblock (\bibinfo{year}{2025}).
\newblock \eprint{2510.27022}.

\bibitem{ParticleDataGroup:2024cfk}
\bibinfo{author}{Navas, S.} \emph{et~al.}
\newblock \bibinfo{journal}{\bibinfo{title}{{Review of particle physics}}}.
\newblock {\emph{\JournalTitle{Phys. Rev. D}}} \textbf{\bibinfo{volume}{110}}, \bibinfo{pages}{030001} (\bibinfo{year}{2024}).

\bibitem{Baudis:2025yva}
\bibinfo{author}{Baudis, L.}
\newblock \bibinfo{journal}{\bibinfo{title}{{Dark matter direct detection: status, results and future plans}}}.
\newblock {\emph{\JournalTitle{J. Phys. Conf. Ser.}}} \textbf{\bibinfo{volume}{3162}}, \bibinfo{pages}{012007} (\bibinfo{year}{2025}).

\bibitem{Cole:2022yzw}
\bibinfo{author}{Cole, P.~S.} \emph{et~al.}
\newblock \bibinfo{journal}{\bibinfo{title}{{Distinguishing environmental effects on binary black hole gravitational waveforms}}}.
\newblock {\emph{\JournalTitle{Nature Astron.}}} \textbf{\bibinfo{volume}{7}}, \bibinfo{pages}{943--950} (\bibinfo{year}{2023}).

\bibitem{McDaniel:2023bju}
\bibinfo{author}{McDaniel, A.} \emph{et~al.}
\newblock \bibinfo{journal}{\bibinfo{title}{{Legacy analysis of dark matter annihilation from the Milky~Way dwarf spheroidal galaxies with 14~years of Fermi-LAT data}}}.
\newblock {\emph{\JournalTitle{Phys. Rev. D}}} \textbf{\bibinfo{volume}{109}}, \bibinfo{pages}{063024} (\bibinfo{year}{2024}).

\bibitem{Lefranc:2015pza}
\bibinfo{author}{Lefranc, V.}, \bibinfo{author}{Moulin, E.}, \bibinfo{author}{Panci, P.} \& \bibinfo{author}{Silk, J.}
\newblock \bibinfo{journal}{\bibinfo{title}{{Prospects for Annihilating Dark Matter in the inner Galactic halo by the Cherenkov Telescope Array}}}.
\newblock {\emph{\JournalTitle{Phys. Rev. D}}} \textbf{\bibinfo{volume}{91}}, \bibinfo{pages}{122003} (\bibinfo{year}{2015}).

\bibitem{BertoneTait:2018}
\bibinfo{author}{Bertone, G.} \& \bibinfo{author}{Tait, T. M.~P.}
\newblock \bibinfo{journal}{\bibinfo{title}{{A new era in the search for dark matter}}}.
\newblock {\emph{\JournalTitle{Nature}}} \textbf{\bibinfo{volume}{562}}, \bibinfo{pages}{51--56} (\bibinfo{year}{2018}).

\bibitem{BertoneCroon:2020GWDM}
\bibinfo{author}{Bertone, G.} \emph{et~al.}
\newblock \bibinfo{journal}{\bibinfo{title}{{Gravitational wave probes of dark matter: challenges and opportunities}}}.
\newblock {\emph{\JournalTitle{SciPost Phys. Core}}} \textbf{\bibinfo{volume}{3}}, \bibinfo{pages}{007} (\bibinfo{year}{2020}).

\bibitem{Bertone:2024DMBHGW}
\bibinfo{author}{Bertone, G.}
\newblock \bibinfo{journal}{\bibinfo{title}{{Dark matter, black holes, and gravitational waves}}}.
\newblock {\emph{\JournalTitle{Nucl. Phys. B}}} \textbf{\bibinfo{volume}{1003}}, \bibinfo{pages}{116487} (\bibinfo{year}{2024}).

\bibitem{Tomaselli:2025jfo}
\bibinfo{author}{Tomaselli, G.~M.}
\newblock \bibinfo{journal}{\bibinfo{title}{{Smooth binary evolution from wide resonances in boson clouds}}}.
\newblock {\emph{\JournalTitle{Phys. Rev. D}}} \textbf{\bibinfo{volume}{112}}, \bibinfo{pages}{063033} (\bibinfo{year}{2025}).

\bibitem{Gondolo:1999ef}
\bibinfo{author}{Gondolo, P.} \& \bibinfo{author}{Silk, J.}
\newblock \bibinfo{journal}{\bibinfo{title}{{Dark matter annihilation at the galactic center}}}.
\newblock {\emph{\JournalTitle{Phys. Rev. Lett.}}} \textbf{\bibinfo{volume}{83}}, \bibinfo{pages}{1719--1722} (\bibinfo{year}{1999}).

\bibitem{Sadeghian:2013laa}
\bibinfo{author}{Sadeghian, L.}, \bibinfo{author}{Ferrer, F.} \& \bibinfo{author}{Will, C.~M.}
\newblock \bibinfo{journal}{\bibinfo{title}{{Dark matter distributions around massive black holes: A general relativistic analysis}}}.
\newblock {\emph{\JournalTitle{Phys. Rev. D}}} \textbf{\bibinfo{volume}{88}}, \bibinfo{pages}{063522} (\bibinfo{year}{2013}).

\bibitem{Bertone:2024wbn}
\bibinfo{author}{Bertone, G.} \emph{et~al.}
\newblock \bibinfo{journal}{\bibinfo{title}{{Toward a realistic description of dark matter overdensities around black holes}}}.
\newblock {\emph{\JournalTitle{Phys. Rev. D}}} \textbf{\bibinfo{volume}{112}}, \bibinfo{pages}{043537} (\bibinfo{year}{2025}).

\bibitem{Caiozzo:2025mye}
\bibinfo{author}{Caiozzo, R.} \emph{et~al.}
\newblock \bibinfo{journal}{\bibinfo{title}{{Dark matter mounds from the collapse of supermassive stars: a general-relativistic analysis}}}.
\newblock (\bibinfo{year}{2025}).
\newblock \eprint{2512.09985}.

\bibitem{Eda:2013gg}
\bibinfo{author}{Eda, K.}, \bibinfo{author}{Itoh, Y.}, \bibinfo{author}{Kuroyanagi, S.} \& \bibinfo{author}{Silk, J.}
\newblock \bibinfo{journal}{\bibinfo{title}{{New Probe of Dark-Matter Properties: Gravitational Waves from an Intermediate-Mass Black Hole Embedded in a Dark-Matter Minispike}}}.
\newblock {\emph{\JournalTitle{Phys. Rev. Lett.}}} \textbf{\bibinfo{volume}{110}}, \bibinfo{pages}{221101} (\bibinfo{year}{2013}).

\bibitem{Kavanagh:2020cfn}
\bibinfo{author}{Kavanagh, B.~J.}, \bibinfo{author}{Nichols, D.~A.}, \bibinfo{author}{Bertone, G.} \& \bibinfo{author}{Gaggero, D.}
\newblock \bibinfo{journal}{\bibinfo{title}{{Detecting dark matter around black holes with gravitational waves: Effects of dark-matter dynamics on the gravitational waveform}}}.
\newblock {\emph{\JournalTitle{Phys. Rev. D}}} \textbf{\bibinfo{volume}{102}}, \bibinfo{pages}{083006} (\bibinfo{year}{2020}).

\bibitem{Coogan:2021uqv}
\bibinfo{author}{Coogan, A.}, \bibinfo{author}{Bertone, G.}, \bibinfo{author}{Gaggero, D.}, \bibinfo{author}{Kavanagh, B.~J.} \& \bibinfo{author}{Nichols, D.~A.}
\newblock \bibinfo{journal}{\bibinfo{title}{{Measuring the dark matter environments of black hole binaries with gravitational waves}}}.
\newblock {\emph{\JournalTitle{Phys. Rev. D}}} \textbf{\bibinfo{volume}{105}}, \bibinfo{pages}{043009} (\bibinfo{year}{2022}).

\bibitem{Green:2024bam}
\bibinfo{author}{Green, A.~M.}
\newblock \bibinfo{journal}{\bibinfo{title}{{Primordial black holes as a dark matter candidate - a brief overview}}}.
\newblock {\emph{\JournalTitle{Nucl. Phys. B}}} \textbf{\bibinfo{volume}{1003}}, \bibinfo{pages}{116494} (\bibinfo{year}{2024}).

\bibitem{Green:2020jor}
\bibinfo{author}{Green, A.~M.} \& \bibinfo{author}{Kavanagh, B.~J.}
\newblock \bibinfo{journal}{\bibinfo{title}{{Primordial Black Holes as a dark matter candidate}}}.
\newblock {\emph{\JournalTitle{J. Phys. G}}} \textbf{\bibinfo{volume}{48}}, \bibinfo{pages}{043001} (\bibinfo{year}{2021}).

\bibitem{Lacki:2010zf}
\bibinfo{author}{Lacki, B.~C.} \& \bibinfo{author}{Beacom, J.~F.}
\newblock \bibinfo{journal}{\bibinfo{title}{{Primordial Black Holes as Dark Matter: Almost All or Almost Nothing}}}.
\newblock {\emph{\JournalTitle{Astrophys. J. Lett.}}} \textbf{\bibinfo{volume}{720}}, \bibinfo{pages}{L67--L71} (\bibinfo{year}{2010}).

\bibitem{MacGibbon:1987my}
\bibinfo{author}{MacGibbon, J.~H.}
\newblock \bibinfo{journal}{\bibinfo{title}{{Can Planck-mass relics of evaporating black holes close the universe?}}}
\newblock {\emph{\JournalTitle{Nature}}} \textbf{\bibinfo{volume}{329}}, \bibinfo{pages}{308--309} (\bibinfo{year}{1987}).

\bibitem{Franciolini:2023osw}
\bibinfo{author}{Franciolini, G.} \& \bibinfo{author}{Pani, P.}
\newblock \bibinfo{journal}{\bibinfo{title}{{Stochastic gravitational-wave background at 3G detectors as a smoking gun for microscopic dark matter relics}}}.
\newblock {\emph{\JournalTitle{Phys. Rev. D}}} \textbf{\bibinfo{volume}{108}}, \bibinfo{pages}{083527} (\bibinfo{year}{2023}).

\bibitem{Aggarwal:2025noe}
\bibinfo{author}{Aggarwal, N.} \emph{et~al.}
\newblock \bibinfo{journal}{\bibinfo{title}{{Challenges and opportunities of gravitational-wave searches above 10 kHz}}}.
\newblock {\emph{\JournalTitle{Living Rev. Rel.}}} \textbf{\bibinfo{volume}{28}}, \bibinfo{pages}{10} (\bibinfo{year}{2025}).

\bibitem{Stephanov:2004wx}
\bibinfo{author}{Stephanov, M.~A.}
\newblock \bibinfo{journal}{\bibinfo{title}{{QCD Phase Diagram and the Critical Point}}}.
\newblock {\emph{\JournalTitle{Prog. Theor. Phys. Suppl.}}} \textbf{\bibinfo{volume}{153}}, \bibinfo{pages}{139--156} (\bibinfo{year}{2004}).

\bibitem{LIGOScientific:2018cki}
\bibinfo{author}{Abbott, B.~P.} \emph{et~al.}
\newblock \bibinfo{journal}{\bibinfo{title}{{GW170817: Measurements of neutron star radii and equation of state}}}.
\newblock {\emph{\JournalTitle{Phys. Rev. Lett.}}} \textbf{\bibinfo{volume}{121}}, \bibinfo{pages}{161101} (\bibinfo{year}{2018}).

\bibitem{Agathos:2019sah}
\bibinfo{author}{Agathos, M.} \emph{et~al.}
\newblock \bibinfo{journal}{\bibinfo{title}{{Inferring Prompt Black-Hole Formation in Neutron Star Mergers from Gravitational-Wave Data}}}.
\newblock {\emph{\JournalTitle{Phys. Rev. D}}} \textbf{\bibinfo{volume}{101}}, \bibinfo{pages}{044006} (\bibinfo{year}{2020}).

\bibitem{Siegel:2022upa}
\bibinfo{author}{Siegel, D.~M.}
\newblock \bibinfo{journal}{\bibinfo{title}{{r-Process nucleosynthesis in gravitational-wave and other explosive astrophysical events}}}.
\newblock {\emph{\JournalTitle{Nature Rev. Phys.}}} \textbf{\bibinfo{volume}{4}}, \bibinfo{pages}{306--318} (\bibinfo{year}{2022}).

\bibitem{Yunes:2022ldq}
\bibinfo{author}{Yunes, N.}, \bibinfo{author}{Miller, M.~C.} \& \bibinfo{author}{Yagi, K.}
\newblock \bibinfo{journal}{\bibinfo{title}{{Gravitational-wave and X-ray probes of the neutron star equation of state}}}.
\newblock {\emph{\JournalTitle{Nature Rev. Phys.}}} \textbf{\bibinfo{volume}{4}}, \bibinfo{pages}{237--246} (\bibinfo{year}{2022}).

\bibitem{Miller:2019cac}
\bibinfo{author}{Miller, M.~C.} \emph{et~al.}
\newblock \bibinfo{journal}{\bibinfo{title}{{PSR J0030+0451 Mass and Radius from $NICER$ Data and Implications for the Properties of Neutron Star Matter}}}.
\newblock {\emph{\JournalTitle{Astrophys. J. Lett.}}} \textbf{\bibinfo{volume}{887}}, \bibinfo{pages}{L24} (\bibinfo{year}{2019}).

\bibitem{Miller:2021qha}
\bibinfo{author}{Miller, M.~C.} \emph{et~al.}
\newblock \bibinfo{journal}{\bibinfo{title}{{The Radius of PSR J0740+6620 from NICER and XMM-Newton Data}}}.
\newblock {\emph{\JournalTitle{Astrophys. J. Lett.}}} \textbf{\bibinfo{volume}{918}}, \bibinfo{pages}{L28} (\bibinfo{year}{2021}).

\bibitem{Fujimoto:2022ohj}
\bibinfo{author}{Fujimoto, Y.}, \bibinfo{author}{Fukushima, K.}, \bibinfo{author}{McLerran, L.~D.} \& \bibinfo{author}{Praszalowicz, M.}
\newblock \bibinfo{journal}{\bibinfo{title}{{Trace Anomaly as Signature of Conformality in Neutron Stars}}}.
\newblock {\emph{\JournalTitle{Phys. Rev. Lett.}}} \textbf{\bibinfo{volume}{129}}, \bibinfo{pages}{252702} (\bibinfo{year}{2022}).

\bibitem{Metag:1992jh}
\bibinfo{author}{Metag, V.}
\newblock \bibinfo{journal}{\bibinfo{title}{{Near threshold particle production: A Probe for resonance matter formation in relativistic heavy ion collisions}}}.
\newblock {\emph{\JournalTitle{Prog. Part. Nucl. Phys.}}} \textbf{\bibinfo{volume}{30}}, \bibinfo{pages}{75--88} (\bibinfo{year}{1993}).

\bibitem{Ribes:2019kno}
\bibinfo{author}{Ribes, P.}, \bibinfo{author}{Ramos, A.}, \bibinfo{author}{Tolos, L.}, \bibinfo{author}{Gonzalez-Boquera, C.} \& \bibinfo{author}{Centelles, M.}
\newblock \bibinfo{journal}{\bibinfo{title}{{Interplay between $\Delta$ Particles and Hyperons in Neutron Stars}}}.
\newblock {\emph{\JournalTitle{Astrophys. J.}}} \textbf{\bibinfo{volume}{883}}, \bibinfo{pages}{168} (\bibinfo{year}{2019}).

\bibitem{Tolos:2024xyi}
\bibinfo{author}{Tolos, L.}
\newblock \bibinfo{journal}{\bibinfo{title}{{Dense Hadronic Matter in Neutron Stars}}}.
\newblock {\emph{\JournalTitle{Acta Phys. Polon. B}}} \textbf{\bibinfo{volume}{55}}, \bibinfo{pages}{5--A1} (\bibinfo{year}{2024}).

\bibitem{McLerran:2018hbz}
\bibinfo{author}{McLerran, L.} \& \bibinfo{author}{Reddy, S.}
\newblock \bibinfo{journal}{\bibinfo{title}{{Quarkyonic Matter and Neutron Stars}}}.
\newblock {\emph{\JournalTitle{Phys. Rev. Lett.}}} \textbf{\bibinfo{volume}{122}}, \bibinfo{pages}{122701} (\bibinfo{year}{2019}).

\bibitem{Zappa:2017xba}
\bibinfo{author}{Zappa, F.}, \bibinfo{author}{Bernuzzi, S.}, \bibinfo{author}{Radice, D.}, \bibinfo{author}{Perego, A.} \& \bibinfo{author}{Dietrich, T.}
\newblock \bibinfo{journal}{\bibinfo{title}{{Gravitational-wave luminosity of binary neutron stars mergers}}}.
\newblock {\emph{\JournalTitle{Phys. Rev. Lett.}}} \textbf{\bibinfo{volume}{120}}, \bibinfo{pages}{111101} (\bibinfo{year}{2018}).

\bibitem{Bernuzzi:2015opx}
\bibinfo{author}{Bernuzzi, S.} \emph{et~al.}
\newblock \bibinfo{journal}{\bibinfo{title}{{How loud are neutron star mergers?}}}
\newblock {\emph{\JournalTitle{Phys. Rev. D}}} \textbf{\bibinfo{volume}{94}}, \bibinfo{pages}{024023} (\bibinfo{year}{2016}).

\bibitem{Hotokezaka:2013iia}
\bibinfo{author}{Hotokezaka, K.} \emph{et~al.}
\newblock \bibinfo{journal}{\bibinfo{title}{{Remnant massive neutron stars of binary neutron star mergers: Evolution process and gravitational waveform}}}.
\newblock {\emph{\JournalTitle{Phys. Rev. D}}} \textbf{\bibinfo{volume}{88}}, \bibinfo{pages}{044026} (\bibinfo{year}{2013}).

\bibitem{Bauswein:2011tp}
\bibinfo{author}{Bauswein, A.} \& \bibinfo{author}{Janka, H.~T.}
\newblock \bibinfo{journal}{\bibinfo{title}{{Measuring neutron-star properties via gravitational waves from binary mergers}}}.
\newblock {\emph{\JournalTitle{Phys. Rev. Lett.}}} \textbf{\bibinfo{volume}{108}}, \bibinfo{pages}{011101} (\bibinfo{year}{2012}).

\bibitem{Dietrich:2016hky}
\bibinfo{author}{Dietrich, T.}, \bibinfo{author}{Ujevic, M.}, \bibinfo{author}{Tichy, W.}, \bibinfo{author}{Bernuzzi, S.} \& \bibinfo{author}{Bruegmann, B.}
\newblock \bibinfo{journal}{\bibinfo{title}{{Gravitational waves and mass ejecta from binary neutron star mergers: Effect of the mass-ratio}}}.
\newblock {\emph{\JournalTitle{Phys. Rev. D}}} \textbf{\bibinfo{volume}{95}}, \bibinfo{pages}{024029} (\bibinfo{year}{2017}).

\bibitem{Breschi:2021xrx}
\bibinfo{author}{Breschi, M.}, \bibinfo{author}{Bernuzzi, S.}, \bibinfo{author}{Godzieba, D.}, \bibinfo{author}{Perego, A.} \& \bibinfo{author}{Radice, D.}
\newblock \bibinfo{journal}{\bibinfo{title}{{Constraints on the Maximum Densities of Neutron Stars from Postmerger Gravitational Waves with Third-Generation Observations}}}.
\newblock {\emph{\JournalTitle{Phys. Rev. Lett.}}} \textbf{\bibinfo{volume}{128}}, \bibinfo{pages}{161102} (\bibinfo{year}{2022}).

\bibitem{Bernuzzi:2015rla}
\bibinfo{author}{Bernuzzi, S.}, \bibinfo{author}{Dietrich, T.} \& \bibinfo{author}{Nagar, A.}
\newblock \bibinfo{journal}{\bibinfo{title}{{Modeling the complete gravitational wave spectrum of neutron star mergers}}}.
\newblock {\emph{\JournalTitle{Phys. Rev. Lett.}}} \textbf{\bibinfo{volume}{115}}, \bibinfo{pages}{091101} (\bibinfo{year}{2015}).

\bibitem{Bauswein:2018bma}
\bibinfo{author}{Bauswein, A.} \emph{et~al.}
\newblock \bibinfo{journal}{\bibinfo{title}{{Identifying a first-order phase transition in neutron star mergers through gravitational waves}}}.
\newblock {\emph{\JournalTitle{Phys. Rev. Lett.}}} \textbf{\bibinfo{volume}{122}}, \bibinfo{pages}{061102} (\bibinfo{year}{2019}).

\bibitem{Radice:2016rys}
\bibinfo{author}{Radice, D.}, \bibinfo{author}{Bernuzzi, S.}, \bibinfo{author}{Del~Pozzo, W.}, \bibinfo{author}{Roberts, L.~F.} \& \bibinfo{author}{Ott, C.~D.}
\newblock \bibinfo{journal}{\bibinfo{title}{{Probing Extreme-Density Matter with Gravitational Wave Observations of Binary Neutron Star Merger Remnants}}}.
\newblock {\emph{\JournalTitle{Astrophys. J. Lett.}}} \textbf{\bibinfo{volume}{842}}, \bibinfo{pages}{L10} (\bibinfo{year}{2017}).

\bibitem{Prakash:2021wpz}
\bibinfo{author}{Prakash, A.} \emph{et~al.}
\newblock \bibinfo{journal}{\bibinfo{title}{{Signatures of deconfined quark phases in binary neutron star mergers}}}.
\newblock {\emph{\JournalTitle{Phys. Rev. D}}} \textbf{\bibinfo{volume}{104}}, \bibinfo{pages}{083029} (\bibinfo{year}{2021}).

\bibitem{Fujimoto:2024ymt}
\bibinfo{author}{Fujimoto, Y.}, \bibinfo{author}{Fukushima, K.}, \bibinfo{author}{Hotokezaka, K.} \& \bibinfo{author}{Kyutoku, K.}
\newblock \bibinfo{journal}{\bibinfo{title}{{Signature of hadron-quark crossover in binary-neutron-star mergers}}}.
\newblock {\emph{\JournalTitle{Phys. Rev. D}}} \textbf{\bibinfo{volume}{111}}, \bibinfo{pages}{063054} (\bibinfo{year}{2025}).

\bibitem{Kapusta:2021ney}
\bibinfo{author}{Kapusta, J.~I.} \& \bibinfo{author}{Welle, T.}
\newblock \bibinfo{journal}{\bibinfo{title}{{Neutron stars with a crossover equation of state}}}.
\newblock {\emph{\JournalTitle{Phys. Rev. C}}} \textbf{\bibinfo{volume}{104}}, \bibinfo{pages}{L012801} (\bibinfo{year}{2021}).

\bibitem{McLerran:2020rnw}
\bibinfo{author}{McLerran, L.}
\newblock \bibinfo{journal}{\bibinfo{title}{{A Pedagogical Discussion of Quarkyonic Matter and Its Implication for Neutron Stars}}}.
\newblock {\emph{\JournalTitle{Acta Phys. Polon. B}}} \textbf{\bibinfo{volume}{51}}, \bibinfo{pages}{1067--1077} (\bibinfo{year}{2020}).

\bibitem{Fujimoto:2022xhv}
\bibinfo{author}{Fujimoto, Y.}, \bibinfo{author}{Fukushima, K.}, \bibinfo{author}{Hotokezaka, K.} \& \bibinfo{author}{Kyutoku, K.}
\newblock \bibinfo{journal}{\bibinfo{title}{{Gravitational Wave Signal for Quark Matter with Realistic Phase Transition}}}.
\newblock {\emph{\JournalTitle{Phys. Rev. Lett.}}} \textbf{\bibinfo{volume}{130}}, \bibinfo{pages}{091404} (\bibinfo{year}{2023}).

\bibitem{Lovato:2022vgq}
\bibinfo{author}{Lovato, A.} \emph{et~al.}
\newblock \bibinfo{journal}{\bibinfo{title}{{Long Range Plan: Dense matter theory for heavy-ion collisions and neutron stars}}}.
\newblock (\bibinfo{year}{2022}).
\newblock \eprint{2211.02224}.

\bibitem{CBM:2016kpk}
\bibinfo{author}{Ablyazimov, T.} \emph{et~al.}
\newblock \bibinfo{journal}{\bibinfo{title}{{Challenges in QCD matter physics --The scientific programme of the Compressed Baryonic Matter experiment at FAIR}}}.
\newblock {\emph{\JournalTitle{Eur. Phys. J. A}}} \textbf{\bibinfo{volume}{53}}, \bibinfo{pages}{60} (\bibinfo{year}{2017}).

\bibitem{Galatyuk:2019lcf}
\bibinfo{author}{Galatyuk, T.}
\newblock \bibinfo{journal}{\bibinfo{title}{{Future facilities for high $\mu_B$ physics}}}.
\newblock {\emph{\JournalTitle{Nucl. Phys. A}}} \textbf{\bibinfo{volume}{982}}, \bibinfo{pages}{163--169} (\bibinfo{year}{2019}).

\bibitem{HADES:2019auv}
\bibinfo{author}{Adamczewski-Musch, J.} \emph{et~al.}
\newblock \bibinfo{journal}{\bibinfo{title}{{Probing dense baryon-rich matter with virtual photons}}}.
\newblock {\emph{\JournalTitle{Nature Phys.}}} \textbf{\bibinfo{volume}{15}}, \bibinfo{pages}{1040--1045} (\bibinfo{year}{2019}).

\bibitem{Geissel:2003lcy}
\bibinfo{author}{Geissel, H.} \emph{et~al.}
\newblock \bibinfo{journal}{\bibinfo{title}{{The Super-FRS project at GSI}}}.
\newblock {\emph{\JournalTitle{Nucl. Instrum. Meth. B}}} \textbf{\bibinfo{volume}{204}}, \bibinfo{pages}{71--85} (\bibinfo{year}{2003}).

\bibitem{Rodriguez-Sanchez:2021zik}
\bibinfo{author}{Rodriguez-Sanchez, J.~L.} \emph{et~al.}
\newblock \bibinfo{journal}{\bibinfo{title}{{Systematic study of {\ensuremath{\Delta}}(1232) resonance excitations using single isobaric charge-exchange reactions induced by medium-mass projectiles of Sn}}}.
\newblock {\emph{\JournalTitle{Phys. Rev. C}}} \textbf{\bibinfo{volume}{106}}, \bibinfo{pages}{014618} (\bibinfo{year}{2022}).

\bibitem{Zhang:2021mks}
\bibinfo{author}{Zhang, J.} \emph{et~al.}
\newblock \bibinfo{journal}{\bibinfo{title}{{First Constraints on Nuclear Coupling of Axionlike Particles from the Binary Neutron Star Gravitational Wave Event GW170817}}}.
\newblock {\emph{\JournalTitle{Phys. Rev. Lett.}}} \textbf{\bibinfo{volume}{127}}, \bibinfo{pages}{161101} (\bibinfo{year}{2021}).

\bibitem{Kumamoto:2024wjd}
\bibinfo{author}{Kumamoto, M.}, \bibinfo{author}{Huang, J.}, \bibinfo{author}{Drischler, C.}, \bibinfo{author}{Baryakhtar, M.} \& \bibinfo{author}{Reddy, S.}
\newblock \bibinfo{journal}{\bibinfo{title}{{Neutron stars with exceptionally light QCD axions}}}.
\newblock {\emph{\JournalTitle{Phys. Rev. D}}} \textbf{\bibinfo{volume}{112}}, \bibinfo{pages}{043008} (\bibinfo{year}{2025}).

\bibitem{IceCube:2026ads}
\bibinfo{author}{Abbasi, R.} \emph{et~al.}
\newblock \bibinfo{journal}{\bibinfo{title}{{Deep Search for Joint Sources of Gravitational Waves and High-Energy Neutrinos with IceCube During the Third Observing Run of LIGO and Virgo}}}.
\newblock (\bibinfo{year}{2026}).
\newblock \eprint{2601.07595}.

\bibitem{IceCube-Gen2:2020qha}
\bibinfo{author}{Aartsen, M.~G.} \emph{et~al.}
\newblock \bibinfo{journal}{\bibinfo{title}{{IceCube-Gen2: the window to the extreme Universe}}}.
\newblock {\emph{\JournalTitle{J. Phys. G}}} \textbf{\bibinfo{volume}{48}}, \bibinfo{pages}{060501} (\bibinfo{year}{2021}).

\bibitem{Planck:2018vyg}
\bibinfo{author}{Aghanim, N.} \emph{et~al.}
\newblock \bibinfo{journal}{\bibinfo{title}{{Planck 2018 results. VI. Cosmological parameters}}}.
\newblock {\emph{\JournalTitle{Astron. Astrophys.}}} \textbf{\bibinfo{volume}{641}}, \bibinfo{pages}{A6} (\bibinfo{year}{2020}).
\newblock \bibinfo{note}{[Erratum: {\it Astron. Astrophys.} {\bf 652}, C4 (2021)]}.

\bibitem{ACT:2025tim}
\bibinfo{author}{Calabrese, E.} \emph{et~al.}
\newblock \bibinfo{journal}{\bibinfo{title}{{The Atacama Cosmology Telescope: DR6 constraints on extended cosmological models}}}.
\newblock {\emph{\JournalTitle{JCAP}}} \textbf{\bibinfo{volume}{11}}, \bibinfo{pages}{063} (\bibinfo{year}{2025}).

\bibitem{SPT-3G:2024qkd}
\bibinfo{author}{Prabhu, K.} \emph{et~al.}
\newblock \bibinfo{journal}{\bibinfo{title}{{Testing the {\ensuremath{\Lambda}}CDM Cosmological Model with Forthcoming Measurements of the Cosmic Microwave Background with SPT-3G}}}.
\newblock {\emph{\JournalTitle{Astrophys. J.}}} \textbf{\bibinfo{volume}{973}}, \bibinfo{pages}{4} (\bibinfo{year}{2024}).

\bibitem{DESI:2024mwx}
\bibinfo{author}{Adame, A.~G.} \emph{et~al.}
\newblock \bibinfo{journal}{\bibinfo{title}{{DESI 2024 VI: cosmological constraints from the measurements of baryon acoustic oscillations}}}.
\newblock {\emph{\JournalTitle{JCAP}}} \textbf{\bibinfo{volume}{02}}, \bibinfo{pages}{021} (\bibinfo{year}{2025}).

\bibitem{DESI:2025zgx}
\bibinfo{author}{Abdul~Karim, M.} \emph{et~al.}
\newblock \bibinfo{journal}{\bibinfo{title}{{DESI DR2 results. II. Measurements of baryon acoustic oscillations and cosmological constraints}}}.
\newblock {\emph{\JournalTitle{Phys. Rev. D}}} \textbf{\bibinfo{volume}{112}}, \bibinfo{pages}{083515} (\bibinfo{year}{2025}).

\bibitem{Riess:2019qba}
\bibinfo{author}{Riess, A.~G.}
\newblock \bibinfo{journal}{\bibinfo{title}{{The Expansion of the Universe is Faster than Expected}}}.
\newblock {\emph{\JournalTitle{Nature Rev. Phys.}}} \textbf{\bibinfo{volume}{2}}, \bibinfo{pages}{10--12} (\bibinfo{year}{2019}).

\bibitem{Efstathiou:2024dvn}
\bibinfo{author}{Efstathiou, G.}
\newblock \bibinfo{journal}{\bibinfo{title}{{Challenges to the {\ensuremath{\Lambda}}CDM cosmology}}}.
\newblock {\emph{\JournalTitle{Phil. Trans. Roy. Soc. Lond. A}}} \textbf{\bibinfo{volume}{383}}, \bibinfo{pages}{20240022} (\bibinfo{year}{2025}).

\bibitem{LIGOScientific:2017adf}
\bibinfo{author}{Abbott, B.~P.} \emph{et~al.}
\newblock \bibinfo{journal}{\bibinfo{title}{{A gravitational-wave standard siren measurement of the Hubble constant}}}.
\newblock {\emph{\JournalTitle{Nature}}} \textbf{\bibinfo{volume}{551}}, \bibinfo{pages}{85--88} (\bibinfo{year}{2017}).

\bibitem{Peebles:1987ek}
\bibinfo{author}{Peebles, P. J.~E.} \& \bibinfo{author}{Ratra, B.}
\newblock \bibinfo{journal}{\bibinfo{title}{{Cosmology with a Time Variable Cosmological Constant}}}.
\newblock {\emph{\JournalTitle{Astrophys. J. Lett.}}} \textbf{\bibinfo{volume}{325}}, \bibinfo{pages}{L17} (\bibinfo{year}{1988}).

\bibitem{Peebles:2002gy}
\bibinfo{author}{Peebles, P. J.~E.} \& \bibinfo{author}{Ratra, B.}
\newblock \bibinfo{journal}{\bibinfo{title}{{The Cosmological Constant and Dark Energy}}}.
\newblock {\emph{\JournalTitle{Rev. Mod. Phys.}}} \textbf{\bibinfo{volume}{75}}, \bibinfo{pages}{559--606} (\bibinfo{year}{2003}).

\bibitem{Wetterich:1987fm}
\bibinfo{author}{Wetterich, C.}
\newblock \bibinfo{journal}{\bibinfo{title}{{Cosmology and the Fate of Dilatation Symmetry}}}.
\newblock {\emph{\JournalTitle{Nucl. Phys.}}} \textbf{\bibinfo{volume}{B302}}, \bibinfo{pages}{668--696} (\bibinfo{year}{1988}).

\bibitem{Wetterich:1994bg}
\bibinfo{author}{Wetterich, C.}
\newblock \bibinfo{journal}{\bibinfo{title}{{The Cosmon model for an asymptotically vanishing time dependent cosmological 'constant'}}}.
\newblock {\emph{\JournalTitle{Astron. Astrophys.}}} \textbf{\bibinfo{volume}{301}}, \bibinfo{pages}{321--328} (\bibinfo{year}{1995}).

\bibitem{Bass:2023ece}
\bibinfo{author}{Bass, S.~D.}
\newblock \bibinfo{journal}{\bibinfo{title}{{The cosmological constant and scale hierarchies with emergent gauge symmetries}}}.
\newblock {\emph{\JournalTitle{Phil. Trans. Roy. Soc. Lond. A}}} \textbf{\bibinfo{volume}{382}}, \bibinfo{pages}{20230092} (\bibinfo{year}{2023}).

\bibitem{Feeney:2018mkj}
\bibinfo{author}{Feeney, S.~M.} \emph{et~al.}
\newblock \bibinfo{journal}{\bibinfo{title}{{Prospects for resolving the Hubble constant tension with standard sirens}}}.
\newblock {\emph{\JournalTitle{Phys. Rev. Lett.}}} \textbf{\bibinfo{volume}{122}}, \bibinfo{pages}{061105} (\bibinfo{year}{2019}).

\bibitem{Feeney:2020kxk}
\bibinfo{author}{Feeney, S.~M.}, \bibinfo{author}{Peiris, H.~V.}, \bibinfo{author}{Nissanke, S.~M.} \& \bibinfo{author}{Mortlock, D.~J.}
\newblock \bibinfo{journal}{\bibinfo{title}{{Prospects for Measuring the Hubble Constant with Neutron-Star{\textendash}Black-Hole Mergers}}}.
\newblock {\emph{\JournalTitle{Phys. Rev. Lett.}}} \textbf{\bibinfo{volume}{126}}, \bibinfo{pages}{171102} (\bibinfo{year}{2021}).

\bibitem{Sarin:2023tgv}
\bibinfo{author}{Sarin, N.} \emph{et~al.}
\newblock \bibinfo{journal}{\bibinfo{title}{{Measuring the nuclear equation of state with neutron star-black hole mergers}}}.
\newblock {\emph{\JournalTitle{Phys. Rev. D}}} \textbf{\bibinfo{volume}{110}}, \bibinfo{pages}{024076} (\bibinfo{year}{2024}).

\bibitem{Weltman:2018zrl}
\bibinfo{author}{Weltman, A.} \emph{et~al.}
\newblock \bibinfo{journal}{\bibinfo{title}{{Fundamental physics with the Square Kilometre Array}}}.
\newblock {\emph{\JournalTitle{Publ. Astron. Soc. Austral.}}} \textbf{\bibinfo{volume}{37}}, \bibinfo{pages}{e002} (\bibinfo{year}{2020}).

\bibitem{Witten:1984rs}
\bibinfo{author}{Witten, E.}
\newblock \bibinfo{journal}{\bibinfo{title}{{Cosmic Separation of Phases}}}.
\newblock {\emph{\JournalTitle{Phys. Rev. D}}} \textbf{\bibinfo{volume}{30}}, \bibinfo{pages}{272--285} (\bibinfo{year}{1984}).

\bibitem{Hogan:1986dsh}
\bibinfo{author}{Hogan, C.~J.}
\newblock \bibinfo{journal}{\bibinfo{title}{{Gravitational radiation from cosmological phase transitions}}}.
\newblock {\emph{\JournalTitle{Mon. Not. Roy. Astron. Soc.}}} \textbf{\bibinfo{volume}{218}}, \bibinfo{pages}{629--636} (\bibinfo{year}{1986}).

\bibitem{Domcke:2024soc}
\bibinfo{author}{Domcke, V.}
\newblock \bibinfo{title}{{Discovery Opportunities with Gravitational Waves -- TASI 2024 Lecture Notes}} (\bibinfo{year}{2024}).
\newblock \eprint{2409.08956}.

\bibitem{DOnofrio:2015gop}
\bibinfo{author}{D'Onofrio, M.} \& \bibinfo{author}{Rummukainen, K.}
\newblock \bibinfo{journal}{\bibinfo{title}{{Standard model cross-over on the lattice}}}.
\newblock {\emph{\JournalTitle{Phys. Rev. D}}} \textbf{\bibinfo{volume}{93}}, \bibinfo{pages}{025003} (\bibinfo{year}{2016}).

\bibitem{Weir:2017wfa}
\bibinfo{author}{Weir, D.~J.}
\newblock \bibinfo{journal}{\bibinfo{title}{{Gravitational waves from a first order electroweak phase transition: a brief review}}}.
\newblock {\emph{\JournalTitle{Phil. Trans. Roy. Soc. Lond. A}}} \textbf{\bibinfo{volume}{376}}, \bibinfo{pages}{20170126} (\bibinfo{year}{2018}).
\newblock \bibinfo{note}{[Erratum: Phil.Trans.Roy.Soc.Lond.A 381, 20230212 (2023)]}.

\bibitem{ATLAS:2025hhd}
\bibinfo{author}{Aad, G.} \emph{et~al.}
\newblock \bibinfo{journal}{\bibinfo{title}{{Study of Higgs boson pair production in the $HH \rightarrow b \overline{b} \gamma\gamma$ final state with 308 fb$^{-1}$ of data collected at $\sqrt{s} = 13$ TeV and 13.6 TeV by the ATLAS experiment}}}.
\newblock (\bibinfo{year}{2025}).
\newblock \eprint{2507.03495}.

\bibitem{CMS:2024awa}
\bibinfo{author}{Hayrapetyan, A.} \emph{et~al.}
\newblock \bibinfo{journal}{\bibinfo{title}{{Constraints on the Higgs boson self-coupling from the combination of single and double Higgs boson production in proton-proton collisions at s=13TeV}}}.
\newblock {\emph{\JournalTitle{Phys. Lett. B}}} \textbf{\bibinfo{volume}{861}}, \bibinfo{pages}{139210} (\bibinfo{year}{2025}).

\bibitem{ATLAS:2024ish}
\bibinfo{author}{Aad, G.} \emph{et~al.}
\newblock \bibinfo{journal}{\bibinfo{title}{{Combination of Searches for Higgs Boson Pair Production in pp Collisions at s=13{\,}{\,}TeV with the ATLAS Detector}}}.
\newblock {\emph{\JournalTitle{Phys. Rev. Lett.}}} \textbf{\bibinfo{volume}{133}}, \bibinfo{pages}{101801} (\bibinfo{year}{2024}).

\bibitem{Stylianou:2023tgg}
\bibinfo{author}{Stylianou, P.} \& \bibinfo{author}{Weiglein, G.}
\newblock \bibinfo{journal}{\bibinfo{title}{{Constraints on the trilinear and quartic Higgs couplings from triple Higgs production at the LHC and beyond}}}.
\newblock {\emph{\JournalTitle{Eur. Phys. J. C}}} \textbf{\bibinfo{volume}{84}}, \bibinfo{pages}{366} (\bibinfo{year}{2024}).

\bibitem{CMS:2025hfp}
\bibinfo{author}{{The ATLAS and CMS Collaborations}}.
\newblock \bibinfo{journal}{\bibinfo{title}{{Highlights of the HL-LHC physics projections by ATLAS and CMS}}}.
\newblock  (\bibinfo{year}{2025}).
\newblock \eprint{2504.00672}.

\bibitem{Bednyakov:2015sca}
\bibinfo{author}{Bednyakov, A.~V.}, \bibinfo{author}{Kniehl, B.~A.}, \bibinfo{author}{Pikelner, A.~F.} \& \bibinfo{author}{Veretin, O.~L.}
\newblock \bibinfo{journal}{\bibinfo{title}{{Stability of the Electroweak Vacuum: Gauge Independence and Advanced Precision}}}.
\newblock {\emph{\JournalTitle{Phys. Rev. Lett.}}} \textbf{\bibinfo{volume}{115}}, \bibinfo{pages}{201802} (\bibinfo{year}{2015}).

\bibitem{Jegerlehner:2013cta}
\bibinfo{author}{Jegerlehner, F.}
\newblock \bibinfo{journal}{\bibinfo{title}{{The Standard model as a low-energy effective theory: what is triggering the Higgs mechanism?}}}
\newblock {\emph{\JournalTitle{Acta Phys. Polon. B}}} \textbf{\bibinfo{volume}{45}}, \bibinfo{pages}{1167} (\bibinfo{year}{2014}).

\bibitem{Degrassi:2012ry}
\bibinfo{author}{Degrassi, G.} \emph{et~al.}
\newblock \bibinfo{journal}{\bibinfo{title}{{Higgs mass and vacuum stability in the Standard Model at NNLO}}}.
\newblock {\emph{\JournalTitle{JHEP}}} \textbf{\bibinfo{volume}{08}}, \bibinfo{pages}{098} (\bibinfo{year}{2012}).

\bibitem{Bahl:2022jnx}
\bibinfo{author}{Bahl, H.}, \bibinfo{author}{Braathen, J.} \& \bibinfo{author}{Weiglein, G.}
\newblock \bibinfo{journal}{\bibinfo{title}{{New Constraints on Extended Higgs Sectors from the Trilinear Higgs Coupling}}}.
\newblock {\emph{\JournalTitle{Phys. Rev. Lett.}}} \textbf{\bibinfo{volume}{129}}, \bibinfo{pages}{231802} (\bibinfo{year}{2022}).

\bibitem{Biekotter:2022kgf}
\bibinfo{author}{Biek{\"o}tter, T.}, \bibinfo{author}{Heinemeyer, S.}, \bibinfo{author}{No, J.~M.}, \bibinfo{author}{Olea-Romacho, M.~O.} \& \bibinfo{author}{Weiglein, G.}
\newblock \bibinfo{journal}{\bibinfo{title}{{The trap in the early Universe: impact on the interplay between gravitational waves and LHC physics in the 2HDM}}}.
\newblock {\emph{\JournalTitle{JCAP}}} \textbf{\bibinfo{volume}{03}}, \bibinfo{pages}{031} (\bibinfo{year}{2023}).

\bibitem{Muratore:2021uqj}
\bibinfo{author}{Muratore, M.}, \bibinfo{author}{Vetrugno, D.}, \bibinfo{author}{Vitale, S.} \& \bibinfo{author}{Hartwig, O.}
\newblock \bibinfo{journal}{\bibinfo{title}{{Time delay interferometry combinations as instrument noise monitors for LISA}}}.
\newblock {\emph{\JournalTitle{Phys. Rev. D}}} \textbf{\bibinfo{volume}{105}}, \bibinfo{pages}{023009} (\bibinfo{year}{2022}).

\bibitem{Muratore:2022nbh}
\bibinfo{author}{Muratore, M.}, \bibinfo{author}{Hartwig, O.}, \bibinfo{author}{Vetrugno, D.}, \bibinfo{author}{Vitale, S.} \& \bibinfo{author}{Weber, W.~J.}
\newblock \bibinfo{journal}{\bibinfo{title}{{Effectiveness of null time-delay interferometry channels as instrument noise monitors in LISA}}}.
\newblock {\emph{\JournalTitle{Phys. Rev. D}}} \textbf{\bibinfo{volume}{107}}, \bibinfo{pages}{082004} (\bibinfo{year}{2023}).

\bibitem{Muratore:2023gxh}
\bibinfo{author}{Muratore, M.}, \bibinfo{author}{Gair, J.} \& \bibinfo{author}{Speri, L.}
\newblock \bibinfo{journal}{\bibinfo{title}{{Impact of the noise knowledge uncertainty for the science exploitation of cosmological and astrophysical stochastic gravitational wave background with LISA}}}.
\newblock {\emph{\JournalTitle{Phys. Rev. D}}} \textbf{\bibinfo{volume}{109}}, \bibinfo{pages}{042001} (\bibinfo{year}{2024}).

\bibitem{Hartwig:2023pft}
\bibinfo{author}{Hartwig, O.}, \bibinfo{author}{Lilley, M.}, \bibinfo{author}{Muratore, M.} \& \bibinfo{author}{Pieroni, M.}
\newblock \bibinfo{journal}{\bibinfo{title}{{Stochastic gravitational wave background reconstruction for a nonequilateral and unequal-noise LISA constellation}}}.
\newblock {\emph{\JournalTitle{Phys. Rev. D}}} \textbf{\bibinfo{volume}{107}}, \bibinfo{pages}{123531} (\bibinfo{year}{2023}).

\bibitem{Alvey:2024uoc}
\bibinfo{author}{Alvey, J.}, \bibinfo{author}{Bhardwaj, U.}, \bibinfo{author}{Domcke, V.}, \bibinfo{author}{Pieroni, M.} \& \bibinfo{author}{Weniger, C.}
\newblock \bibinfo{journal}{\bibinfo{title}{{Leveraging time-dependent instrumental noise for the LISA stochastic gravitational wave background analysis}}}.
\newblock {\emph{\JournalTitle{Phys. Rev. D}}} \textbf{\bibinfo{volume}{111}}, \bibinfo{pages}{102006} (\bibinfo{year}{2025}).

\bibitem{Pozzoli:2023lgz}
\bibinfo{author}{Pozzoli, F.}, \bibinfo{author}{Buscicchio, R.}, \bibinfo{author}{Moore, C.~J.}, \bibinfo{author}{Haardt, F.} \& \bibinfo{author}{Sesana, A.}
\newblock \bibinfo{journal}{\bibinfo{title}{{Weakly parametric approach to stochastic background inference in LISA}}}.
\newblock {\emph{\JournalTitle{Phys. Rev. D}}} \textbf{\bibinfo{volume}{109}}, \bibinfo{pages}{083029} (\bibinfo{year}{2024}).

\bibitem{Caprini:2024hue}
\bibinfo{author}{Caprini, C.} \emph{et~al.}
\newblock \bibinfo{journal}{\bibinfo{title}{{Gravitational waves from first-order phase transitions in LISA: reconstruction pipeline and physics interpretation}}}.
\newblock {\emph{\JournalTitle{JCAP}}} \textbf{\bibinfo{volume}{10}}, \bibinfo{pages}{020} (\bibinfo{year}{2024}).

\bibitem{Gong:2021gvw}
\bibinfo{author}{Gong, Y.}, \bibinfo{author}{Luo, J.} \& \bibinfo{author}{Wang, B.}
\newblock \bibinfo{journal}{\bibinfo{title}{{Concepts and status of Chinese space gravitational wave detection projects}}}.
\newblock {\emph{\JournalTitle{Nature Astron.}}} \textbf{\bibinfo{volume}{5}}, \bibinfo{pages}{881--889} (\bibinfo{year}{2021}).

\bibitem{TianQin:2015yph}
\bibinfo{author}{Luo, J.} \emph{et~al.}
\newblock \bibinfo{journal}{\bibinfo{title}{{TianQin: a space-borne gravitational wave detector}}}.
\newblock {\emph{\JournalTitle{Class. Quant. Grav.}}} \textbf{\bibinfo{volume}{33}}, \bibinfo{pages}{035010} (\bibinfo{year}{2016}).

\bibitem{Luo:2025sos}
\bibinfo{author}{Luo, J.} \emph{et~al.}
\newblock \bibinfo{journal}{\bibinfo{title}{{Progress of the TianQin project}}}.
\newblock {\emph{\JournalTitle{Class. Quant. Grav.}}} \textbf{\bibinfo{volume}{42}}, \bibinfo{pages}{173001} (\bibinfo{year}{2025}).

\bibitem{Ruan:2018tsw}
\bibinfo{author}{Ruan, W.-H.}, \bibinfo{author}{Guo, Z.-K.}, \bibinfo{author}{Cai, R.-G.} \& \bibinfo{author}{Zhang, Y.-Z.}
\newblock \bibinfo{journal}{\bibinfo{title}{{Taiji program: Gravitational-wave sources}}}.
\newblock {\emph{\JournalTitle{Int. J. Mod. Phys. A}}} \textbf{\bibinfo{volume}{35}}, \bibinfo{pages}{2050075} (\bibinfo{year}{2020}).

\bibitem{LIGOScientific:2025kry}
\bibinfo{journal}{\bibinfo{title}{{Cosmological and High Energy Physics implications from gravitational-wave background searches in LIGO-Virgo-KAGRA's O1-O4a runs}}}.
\newblock (\bibinfo{year}{2025}).
\newblock \eprint{2510.26848}.

\bibitem{Ackermann:2023gmd}
\bibinfo{author}{Ackermann, M.} \& \bibinfo{author}{Helbing, K.}
\newblock \bibinfo{journal}{\bibinfo{title}{{Searches for beyond-standard-model physics with astroparticle physics instruments}}}.
\newblock {\emph{\JournalTitle{Phil. Trans. Roy. Soc. Lond. A}}} \textbf{\bibinfo{volume}{382}}, \bibinfo{pages}{20230082} (\bibinfo{year}{2023}).

\bibitem{ATLAS:2022vkf}
\bibinfo{author}{Aad, G.} \emph{et~al.}
\newblock \bibinfo{journal}{\bibinfo{title}{{A detailed map of Higgs boson interactions by the ATLAS experiment ten years after the discovery}}}.
\newblock {\emph{\JournalTitle{Nature}}} \textbf{\bibinfo{volume}{607}}, \bibinfo{pages}{52--59} (\bibinfo{year}{2022}).
\newblock \bibinfo{note}{[Erratum: {\it Nature} {\bf 612}, E24 (2022)]}.

\bibitem{LinearColliderVision:2025hlt}
\bibinfo{author}{Abramowicz, H.} \emph{et~al.}
\newblock \bibinfo{journal}{\bibinfo{title}{{A Linear Collider Vision for the Future of Particle Physics}}}.
\newblock (\bibinfo{year}{2025}).
\newblock \eprint{2503.19983}.

\bibitem{Weinberg:1979sa}
\bibinfo{author}{Weinberg, S.}
\newblock \bibinfo{journal}{\bibinfo{title}{{Baryon and Lepton Nonconserving Processes}}}.
\newblock {\emph{\JournalTitle{Phys. Rev. Lett.}}} \textbf{\bibinfo{volume}{43}}, \bibinfo{pages}{1566--1570} (\bibinfo{year}{1979}).

\bibitem{Altarelli:2013tya}
\bibinfo{author}{Altarelli, G.}
\newblock \bibinfo{journal}{\bibinfo{title}{{Collider Physics within the Standard Model: a Primer}}}.
\newblock (\bibinfo{year}{2013}).
\newblock \eprint{1303.2842}.

\bibitem{KATRIN:2024cdt}
\bibinfo{author}{Aker, M.} \emph{et~al.}
\newblock \bibinfo{journal}{\bibinfo{title}{{Direct neutrino-mass measurement based on 259 days of KATRIN data}}}.
\newblock {\emph{\JournalTitle{Science}}} \textbf{\bibinfo{volume}{388}}, \bibinfo{pages}{adq9592} (\bibinfo{year}{2025}).

\bibitem{Bass:2020egf}
\bibinfo{author}{Bass, S.~D.} \& \bibinfo{author}{Krzysiak, J.}
\newblock \bibinfo{journal}{\bibinfo{title}{{Vacuum energy with mass generation and Higgs bosons}}}.
\newblock {\emph{\JournalTitle{Phys. Lett. B}}} \textbf{\bibinfo{volume}{803}}, \bibinfo{pages}{135351} (\bibinfo{year}{2020}).

\bibitem{Bass:2021wxv}
\bibinfo{author}{Bass, S.~D.}
\newblock \bibinfo{journal}{\bibinfo{title}{{Emergent gauge symmetries: making symmetry as well as breaking it}}}.
\newblock {\emph{\JournalTitle{Phil. Trans. A. Math. Phys. Eng. Sci.}}} \textbf{\bibinfo{volume}{380}}, \bibinfo{pages}{20210059} (\bibinfo{year}{2021}).

\bibitem{Bjorken:2001pe}
\bibinfo{author}{Bjorken, J.}
\newblock \bibinfo{title}{{Emergent gauge bosons}}.
\newblock In \emph{\bibinfo{booktitle}{{Proceedings to the workshops: What comes beyond the standard model 2000, 2001. Volume 1}}} (\bibinfo{year}{2001}).
\newblock \eprint{hep-th/0111196}.

\bibitem{Baumann:2008bn}
\bibinfo{author}{Baumann, D.} \& \bibinfo{author}{Peiris, H.~V.}
\newblock \bibinfo{journal}{\bibinfo{title}{{Cosmological Inflation: Theory and Observations}}}.
\newblock {\emph{\JournalTitle{Adv. Sci. Lett.}}} \textbf{\bibinfo{volume}{2}}, \bibinfo{pages}{105--120} (\bibinfo{year}{2009}).

\bibitem{Komatsu:2022nvu}
\bibinfo{author}{Komatsu, E.}
\newblock \bibinfo{journal}{\bibinfo{title}{{New physics from the polarized light of the cosmic microwave background}}}.
\newblock {\emph{\JournalTitle{Nature Rev. Phys.}}} \textbf{\bibinfo{volume}{4}}, \bibinfo{pages}{452--469} (\bibinfo{year}{2022}).

\bibitem{Lyth:1996im}
\bibinfo{author}{Lyth, D.~H.}
\newblock \bibinfo{journal}{\bibinfo{title}{{What would we learn by detecting a gravitational wave signal in the cosmic microwave background anisotropy?}}}
\newblock {\emph{\JournalTitle{Phys. Rev. Lett.}}} \textbf{\bibinfo{volume}{78}}, \bibinfo{pages}{1861--1863} (\bibinfo{year}{1997}).

\bibitem{Tristram:2021tvh}
\bibinfo{author}{Tristram, M.} \emph{et~al.}
\newblock \bibinfo{journal}{\bibinfo{title}{{Improved limits on the tensor-to-scalar ratio using BICEP and Planck data}}}.
\newblock {\emph{\JournalTitle{Phys. Rev. D}}} \textbf{\bibinfo{volume}{105}}, \bibinfo{pages}{083524} (\bibinfo{year}{2022}).

\bibitem{SimonsObservatory:2024gol}
\bibinfo{author}{Hertig, E.} \emph{et~al.}
\newblock \bibinfo{journal}{\bibinfo{title}{{The Simons Observatory: Combining cross-spectral foreground cleaning with multitracer B-mode delensing for improved constraints on inflation}}}.
\newblock {\emph{\JournalTitle{Phys. Rev. D}}} \textbf{\bibinfo{volume}{110}}, \bibinfo{pages}{043532} (\bibinfo{year}{2024}).

\bibitem{LiteBIRD:2022cnt}
\bibinfo{author}{Allys, E.} \emph{et~al.}
\newblock \bibinfo{journal}{\bibinfo{title}{{Probing Cosmic Inflation with the LiteBIRD Cosmic Microwave Background Polarization Survey}}}.
\newblock {\emph{\JournalTitle{PTEP}}} \textbf{\bibinfo{volume}{2023}}, \bibinfo{pages}{042F01} (\bibinfo{year}{2023}).

\bibitem{LiteBIRD:2023iei}
\bibinfo{author}{Fuskeland, U.} \emph{et~al.}
\newblock \bibinfo{journal}{\bibinfo{title}{{Tensor-to-scalar ratio forecasts for extended LiteBIRD frequency configurations}}}.
\newblock {\emph{\JournalTitle{Astron. Astrophys.}}} \textbf{\bibinfo{volume}{676}}, \bibinfo{pages}{A42} (\bibinfo{year}{2023}).

\bibitem{Ghoshal:2025ejg}
\bibinfo{author}{Ghoshal, A.}, \bibinfo{author}{Koz{\'o}w, P.}, \bibinfo{author}{Olechowski, M.} \& \bibinfo{author}{Pokorski, S.}
\newblock \bibinfo{journal}{\bibinfo{title}{{CMB observables and reheat temperature as a window to models of inflation and freeze-in dark matter production}}}.
\newblock (\bibinfo{year}{2025}).
\newblock \eprint{2510.27587}.

\bibitem{Maggiore:1999vm}
\bibinfo{author}{Maggiore, M.}
\newblock \bibinfo{journal}{\bibinfo{title}{{Gravitational wave experiments and early universe cosmology}}}.
\newblock {\emph{\JournalTitle{Phys. Rept.}}} \textbf{\bibinfo{volume}{331}}, \bibinfo{pages}{283--367} (\bibinfo{year}{2000}).

\bibitem{Bartolo:2016ami}
\bibinfo{author}{Bartolo, N.} \emph{et~al.}
\newblock \bibinfo{journal}{\bibinfo{title}{{Science with the space-based interferometer LISA. IV: Probing inflation with gravitational waves}}}.
\newblock {\emph{\JournalTitle{JCAP}}} \textbf{\bibinfo{volume}{12}}, \bibinfo{pages}{026} (\bibinfo{year}{2016}).

\bibitem{Ricciardone:2016ddg}
\bibinfo{author}{Ricciardone, A.}
\newblock \bibinfo{journal}{\bibinfo{title}{{Primordial Gravitational Waves with LISA}}}.
\newblock {\emph{\JournalTitle{J. Phys. Conf. Ser.}}} \textbf{\bibinfo{volume}{840}}, \bibinfo{pages}{012030} (\bibinfo{year}{2017}).

\bibitem{Guzzetti:2016mkm}
\bibinfo{author}{Guzzetti, M.~C.}, \bibinfo{author}{Bartolo, N.}, \bibinfo{author}{Liguori, M.} \& \bibinfo{author}{Matarrese, S.}
\newblock \bibinfo{journal}{\bibinfo{title}{{Gravitational waves from inflation}}}.
\newblock {\emph{\JournalTitle{Riv. Nuovo Cim.}}} \textbf{\bibinfo{volume}{39}}, \bibinfo{pages}{399--495} (\bibinfo{year}{2016}).

\bibitem{Caprini:2018mtu}
\bibinfo{author}{Caprini, C.} \& \bibinfo{author}{Figueroa, D.~G.}
\newblock \bibinfo{journal}{\bibinfo{title}{{Cosmological Backgrounds of Gravitational Waves}}}.
\newblock {\emph{\JournalTitle{Class. Quant. Grav.}}} \textbf{\bibinfo{volume}{35}}, \bibinfo{pages}{163001} (\bibinfo{year}{2018}).

\bibitem{Maggiore:2024cwf}
\bibinfo{author}{Maggiore, M.}, \bibinfo{author}{Iacovelli, F.}, \bibinfo{author}{Belgacem, E.}, \bibinfo{author}{Mancarella, M.} \& \bibinfo{author}{Muttoni, N.}
\newblock \bibinfo{journal}{\bibinfo{title}{{Comparison of global networks of third-generation gravitational-wave detectors}}}.
\newblock {\emph{\JournalTitle{Class. Quant. Grav.}}} \textbf{\bibinfo{volume}{42}}, \bibinfo{pages}{215004} (\bibinfo{year}{2025}).

\bibitem{Hild:2010id}
\bibinfo{author}{Hild, S.} \emph{et~al.}
\newblock \bibinfo{journal}{\bibinfo{title}{{Sensitivity Studies for Third-Generation Gravitational Wave Observatories}}}.
\newblock {\emph{\JournalTitle{Class. Quant. Grav.}}} \textbf{\bibinfo{volume}{28}}, \bibinfo{pages}{094013} (\bibinfo{year}{2011}).

\bibitem{Reitze:2019dyk}
\bibinfo{author}{Reitze, D.} \emph{et~al.}
\newblock \bibinfo{journal}{\bibinfo{title}{{The US Program in Ground-Based Gravitational Wave Science: Contribution from the LIGO Laboratory}}}.
\newblock {\emph{\JournalTitle{Bull. Am. Astron. Soc.}}} \textbf{\bibinfo{volume}{51}}, \bibinfo{pages}{141} (\bibinfo{year}{2019}).

\bibitem{Hall:2020dps}
\bibinfo{author}{Hall, E.~D.} \emph{et~al.}
\newblock \bibinfo{journal}{\bibinfo{title}{{Gravitational-wave physics with Cosmic Explorer: Limits to low-frequency sensitivity}}}.
\newblock {\emph{\JournalTitle{Phys. Rev. D}}} \textbf{\bibinfo{volume}{103}}, \bibinfo{pages}{122004} (\bibinfo{year}{2021}).

\bibitem{Bhardwaj:2023xph}
\bibinfo{author}{Bhardwaj, U.}, \bibinfo{author}{Alvey, J.}, \bibinfo{author}{Miller, B.~K.}, \bibinfo{author}{Nissanke, S.} \& \bibinfo{author}{Weniger, C.}
\newblock \bibinfo{journal}{\bibinfo{title}{{Sequential simulation-based inference for gravitational wave signals}}}.
\newblock {\emph{\JournalTitle{Phys. Rev. D}}} \textbf{\bibinfo{volume}{108}}, \bibinfo{pages}{042004} (\bibinfo{year}{2023}).

\bibitem{Alvey:2023npw}
\bibinfo{author}{Alvey, J.}, \bibinfo{author}{Bhardwaj, U.}, \bibinfo{author}{Domcke, V.}, \bibinfo{author}{Pieroni, M.} \& \bibinfo{author}{Weniger, C.}
\newblock \bibinfo{journal}{\bibinfo{title}{{Simulation-based inference for stochastic gravitational wave background data analysis}}}.
\newblock {\emph{\JournalTitle{Phys. Rev. D}}} \textbf{\bibinfo{volume}{109}}, \bibinfo{pages}{083008} (\bibinfo{year}{2024}).

\bibitem{Bass:2023hoi}
\bibinfo{author}{Bass, S.~D.} \& \bibinfo{author}{Doser, M.}
\newblock \bibinfo{journal}{\bibinfo{title}{{Quantum sensing for particle physics}}}.
\newblock {\emph{\JournalTitle{Nature Rev. Phys.}}} \textbf{\bibinfo{volume}{6}}, \bibinfo{pages}{329--339} (\bibinfo{year}{2024}).

\bibitem{TVLBAIProto:2025wyn}
\bibinfo{author}{Bala{\v{z}}, A.} \emph{et~al.}
\newblock \bibinfo{journal}{\bibinfo{title}{{Long-Baseline Atom Interferometry}}}.
\newblock (\bibinfo{year}{2025}).
\newblock \eprint{2503.21366}.

\bibitem{Alonso:2022oot}
\bibinfo{author}{Alonso, I.} \emph{et~al.}
\newblock \bibinfo{journal}{\bibinfo{title}{{Cold atoms in space: community workshop summary and proposed road-map}}}.
\newblock {\emph{\JournalTitle{EPJ Quant. Technol.}}} \textbf{\bibinfo{volume}{9}}, \bibinfo{pages}{30} (\bibinfo{year}{2022}).

\bibitem{Schmieden:2023fzn}
\bibinfo{author}{Schmieden, K.} \& \bibinfo{author}{Schott, M.}
\newblock \bibinfo{journal}{\bibinfo{title}{{A Global Network of Cavities to Search for Gravitational Waves (GravNet): A novel scheme to hunt gravitational waves signatures from the early universe}}}.
\newblock {\emph{\JournalTitle{PoS}}} \textbf{\bibinfo{volume}{EPS-HEP2023}}, \bibinfo{pages}{102} (\bibinfo{year}{2024}).
\newblock \eprint{2308.11497}.

\bibitem{Domcke:2022rgu}
\bibinfo{author}{Domcke, V.}, \bibinfo{author}{Garcia-Cely, C.} \& \bibinfo{author}{Rodd, N.~L.}
\newblock \bibinfo{journal}{\bibinfo{title}{{Novel Search for High-Frequency Gravitational Waves with Low-Mass Axion Haloscopes}}}.
\newblock {\emph{\JournalTitle{Phys. Rev. Lett.}}} \textbf{\bibinfo{volume}{129}}, \bibinfo{pages}{041101} (\bibinfo{year}{2022}).

\bibitem{Capdevilla:2025omb}
\bibinfo{author}{Capdevilla, R.}, \bibinfo{author}{Harnik, R.}, \bibinfo{author}{Kim, T.} \& \bibinfo{author}{Krokotsch, T.}
\newblock \bibinfo{journal}{\bibinfo{title}{{High-frequency gravitational wave detection by the BREAD experiment}}}.
\newblock {\emph{\JournalTitle{Phys. Rev. D}}} \textbf{\bibinfo{volume}{112}}, \bibinfo{pages}{035031} (\bibinfo{year}{2025}).

\bibitem{Kim:2025izt}
\bibinfo{author}{Kim, Y.} \emph{et~al.}
\newblock \bibinfo{journal}{\bibinfo{title}{{Search for high-frequency gravitational waves via re-analysis of cavity axion data}}}.
\newblock (\bibinfo{year}{2025}).
\newblock \eprint{2511.17817}.

\bibitem{Jungkind:2025oqm}
\bibinfo{author}{Jungkind, C.~M.}, \bibinfo{author}{Seymour, B.~C.}, \bibinfo{author}{Laeuger, A.} \& \bibinfo{author}{Chen, Y.}
\newblock \bibinfo{journal}{\bibinfo{title}{{Prospects for high-frequency gravitational-wave detection with GEO600}}}.
\newblock {\emph{\JournalTitle{Phys. Rev. D}}} \textbf{\bibinfo{volume}{113}}, \bibinfo{pages}{024057} (\bibinfo{year}{2026}).

\end{thebibliography}

\end{document}